\documentclass[acmtog, balance = false]{acmart}

\acmArticle{181}
\acmJournal{TOG}
\setcitestyle{square}

\usepackage{wrapfig}
\usepackage{nicefrac}
\usepackage{mathtools}
\usepackage{graphicx}
\usepackage{booktabs} 
\usepackage{combelow} 
\usepackage[ruled,vlined,linesnumbered]{algorithm2e}
\usepackage{tabularx} 
\usepackage{colortbl} 
\usepackage{cancel}
\usepackage{makecell} 
\usepackage{enumitem}
\usepackage{manfnt} 
\usepackage[nameinlink]{cleveref}

\usepackage{ccicons}
\usepackage{amsmath}
\usepackage{amsfonts}
\usepackage{amsbsy}

\usepackage{listings} 
\usepackage{courier}
\definecolor{mygreen}{rgb}{0,0.6,0}
\definecolor{mygray}{rgb}{0.5,0.5,0.5}
\definecolor{mymauve}{rgb}{0.58,0,0.82}
\definecolor{superlightgray}{RGB}{240,240,240}
\definecolor{darkBlue}{RGB}{0, 94, 184}
\definecolor{derekBlue}{RGB}{144,210,236}
\definecolor{derekTableBlue}{RGB}{189,235,252}
\definecolor{iglGreen}{RGB}{153,203,67}
\definecolor{coralRed}{RGB}{250,114,104}
\definecolor{gray}{RGB}{180,180,180}
\definecolor{orange}{RGB}{255,165,0}
\definecolor{TechnionBlue}{RGB}{8,33,78}
\definecolor{Purple}{RGB}{137, 99, 198}
\definecolor{lightgray}{gray}{0.65}

\newcommand{\update}[1]{#1}
\newcommand{\updatetwo}[1]{#1}

\SetCommentSty{commentFont}
\SetNlSty{textbf}{}{.}

\newcommand*{\refsec}[1]{%
  \begingroup
    \def\sectionautorefname{Sec.}%
    \def\subsectionautorefname{Sec.}%
    \def\subsubsectionautorefname{Sec.}%
    \autoref{sec:#1}%
  \endgroup
}
\newcommand*{\refequ}[1]{%
  \begingroup
    \def\equationautorefname{Eq.}
    \autoref{equ:#1}%
  \endgroup
}

\newcommand*{\reffig}[1]{%
  \begingroup
    \def\figureautorefname{Fig.}%
    \autoref{fig:#1}%
  \endgroup
}
\newcommand*{\reffignum}[1]{%
  \begingroup
    \def\figureautorefname{}%
    \autoref{fig:#1}%
  \endgroup
}
\newcommand*{\reftab}[1]{%
  \begingroup
    \def\tableautorefname{Tab.}%
    \autoref{tab:#1}%
  \endgroup
}

\newcommand{\R}{\mathbb{R}}
\newcommand{\M}{\mathcal{M}}
\renewcommand{\P}{\mathcal{P}}
\newcommand{\n}{\hat{\vn}}

\newcommand{\Qa}{Q^A}

\newcommand{\LT}{[L98+T20]\ }
\newcommand{\GLT}{[G97+L98+T20]\ }

\newcommand{\vecFont}[1]{\mathbf{#1}}

\def\vb{{\vecFont{b}}}

\def\vn{{\vecFont{n}}}

\def\vp{{\vecFont{p}}}

\def\vs{{\vecFont{s}}}
\def\vt{{\vecFont{t}}}

\def\vv{{\vecFont{v}}}

\def\vx{{\vecFont{x}}}

\newcommand{\matFont}[1]{\mathbf{#1}}
\def\mA{{\matFont{A}}}

\begin{document}
\title{Simplifying Textured Triangle Meshes in the Wild
} 

\author{Hsueh-Ti Derek Liu}
\affiliation{%
  \institution{Roblox}
  \country{Canada}
  }
\email{hsuehtil@gmail.com}

\author{Xiaoting Zhang}
\affiliation{%
  \institution{Roblox}
  \country{USA}
  }
\email{xzhang@roblox.com}

\author{Cem Yuksel}
\affiliation{%
  \institution{University of Utah \& Roblox}
  \country{USA}
  }
\email{cem@cemyuksel.com} 





\setcopyright{acmlicensed}
\acmJournal{TOG}
\acmYear{2025} \acmVolume{44} \acmNumber{6} \acmArticle{} \acmMonth{12}\acmDOI{10.1145/3763277}

\begin{teaserfigure}
    \includegraphics[width=\linewidth]{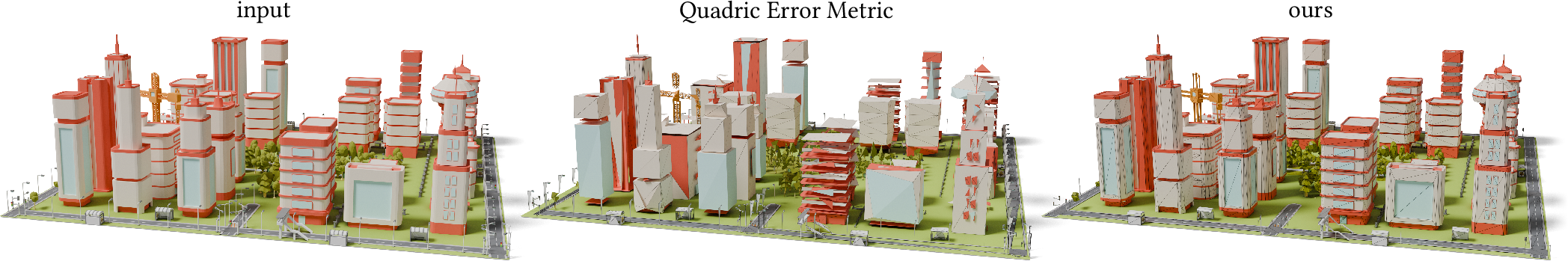}
    \caption{Given an artist-created city scene (\ccby\ \emph{mertkilic}) in the wild, we simplify it down to 10\% of its original resolution. Our method (right) better preserves geometrically significant components (e.g., buildings) compared to the quadric error simplification \cite{QEMWithTexture}, implemented in \cite{meshlab}, (middle) which prioritizes small components (e.g., streetlamps).}
    \label{fig:teaser}
  \end{teaserfigure} 

\begin{abstract}
    This paper introduces a method for simplifying textured surface triangle meshes in the wild while maintaining high visual quality. 
    While previous methods achieve excellent results on \emph{manifold} meshes by using the quadric error metric, they struggle to produce high-quality outputs for meshes in the wild, which typically contain \emph{non-manifold} elements and multiple connected components. 
    %
    %
    In this work, we propose a method for simplifying these ``wild'' textured triangle meshes.
    We formulate mesh simplification as a problem of decimating \emph{simplicial 2-complexes} to handle multiple non-manifold mesh components collectively. 
    Building on the success of quadric error simplification, we iteratively collapse 1-simplices (vertex pairs). Our approach employs a modified quadric error that converges to the original quadric error metric for watertight manifold meshes, while significantly improving the results on wild meshes.
    For textures, instead of following existing strategies to preserve UVs, we adopt a novel perspective which focuses on computing mesh correspondences throughout the decimation, independent of the UV layout.
    This combination yields a textured mesh simplification system that is capable of handling arbitrary triangle meshes, achieving to high-quality results on wild inputs without sacrificing the excellent performance on clean inputs.
    Our method guarantees to avoid common problems in textured mesh simplification, including the prevalent problem of \emph{texture bleeding}.
    We extensively evaluate our method on multiple datasets, showing improvements over prior techniques through qualitative, quantitative, and user study evaluations.
\end{abstract}

\ccsdesc[500]{Computing methodologies~Mesh geometry models}

\keywords{geometry processing, mesh simplification}

\setcopyright{acmlicensed}
\acmJournal{TOG}
\acmYear{2025} \acmVolume{44} \acmNumber{6} \acmArticle{} \acmMonth{12}\acmDOI{10.1145/3763277}

\maketitle
\section{Introduction}\label{sec:introduction}
\begin{figure}
    \begin{center}
    \includegraphics[width=1\linewidth]{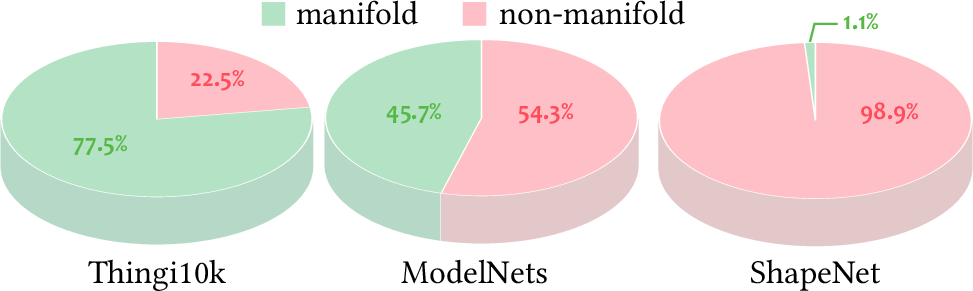}
    \end{center}
    \vspace{-3pt}
    \caption{\update{Many meshes available in online repositories are non-manifold. For instance, Thingi10k \cite{Thingi10K} (a dataset consisting of 3D-printable shapes) includes 22.5\% non-manifold shapes, ModelNet \cite{WuSKYZTX15} (a dataset of CAD models) contains 54.3\% non-manifold objects, and more severely 98.9\% of the meshes in the ShapeNet dataset \cite{shapenet2015} (a widely used dataset for machine learning) are non-manifold.}}
    \label{fig:pie_charts}
\end{figure}

Mesh simplification creates different levels of detail (LOD) for 3D objects while maintaining their visual fidelity.
\begin{figure}
    \begin{center}
    \includegraphics[width=1\linewidth]{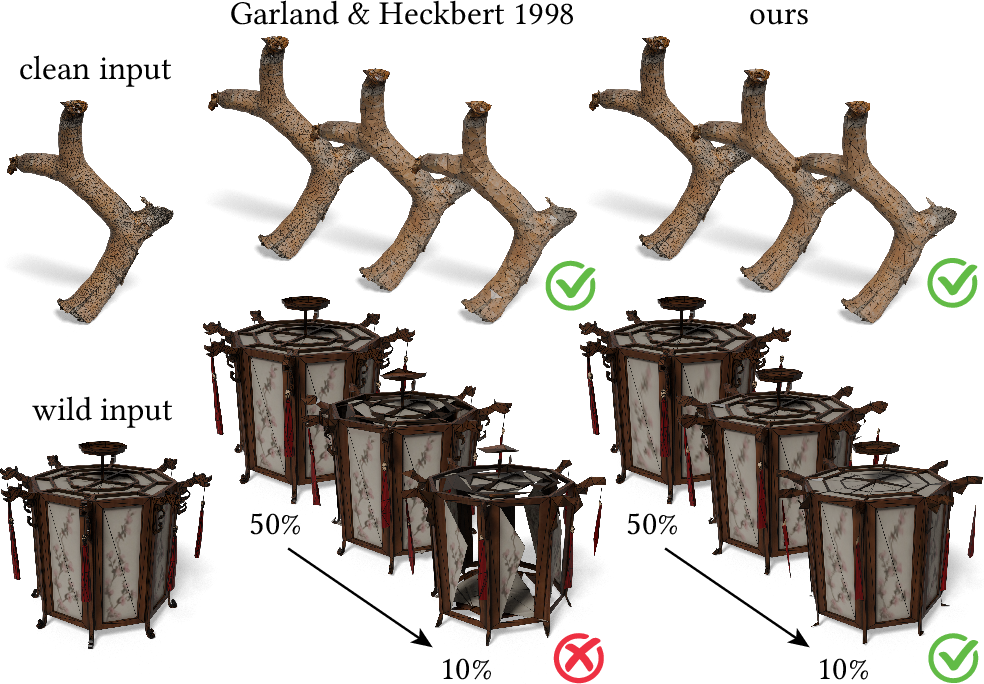}
    \end{center}
    \caption{Comparison between our textured mesh simplification method and a representative prior technique \cite{QEMWithTexture} (using the implementation from \cite{meshlab}). On single component and manifold inputs, both methods produce excellent results (compare baseline, top center, vs. ours, top right). However, for challenging ``wild'' inputs, often characterized by non-manifold and multiple components geometry, the baseline approach frequently yields unsatisfactory results (bottom center). In contrast, our method preserves visual fidelity and geometric structure more effectively on these wild meshes (bottom right).}
    \label{fig:manifold_vs_nonmanifold}
\end{figure}
It plays a critical role in interactive graphics to achieve target performance in rendering and simulation across various hardware platforms.
Its importance has been motivating decades of development with a plethora of solutions, such as the widely used QEM by \citet{GarlandH97}.
While these solutions produce high-quality results for manifold surface inputs, we demonstrate that they often fail to generate acceptable results \update{on meshes} obtained from online repositories (see \reffig{teaser}).
This issue is caused by the discrepancy between existing problem formulations and real-world data characteristics. 
Existing LOD techniques often assume the input mesh is manifold \cite{dey1999topology}.
This was a realistic assumption when, for instance during the 1990s, the goal was to simplify meshes generated by isosurfacing methods, such as marching cubes \cite{LorensenC87}. 
However, recent advancements in 3D modeling have populated online repositories with a rich amount of manually modeled shapes. 
In contrast to meshes reconstructed from isosurfacing, the majority of them are non-manifold (e.g., 98.9\% on ShapeNet \cite{shapenet2015} are non-manifold, see \reffig{pie_charts}). 
%
This prevalence of non-manifold meshes underscores the need to revisit mesh simplification and its problem formulation to better suit contemporary datasets.
%

A desired mesh simplification method must be \emph{performant} and \emph{robust} to all types of geometric artifacts, and can \emph{preserve surface attributes} (e.g., textures).
In real-time applications such as virtual reality and video games, these (defective) meshes appear frequently during game time from user interactions (i.e., engineers collaboratively model 3D buildings).
Being able to efficiently and robustly simplify all the newly created 3D content on the fly while preserving attributes is critical to achieving target performance.
Such requirements also exclude expensive alternatives that do not preserve attributes (see \reffig{tetwild_textured_sim}), such as repairing defective meshes into manifolds and performing simplification afterward, which may take minutes to hours.

\begin{figure}
    \begin{center}
    \includegraphics[width=1\linewidth]{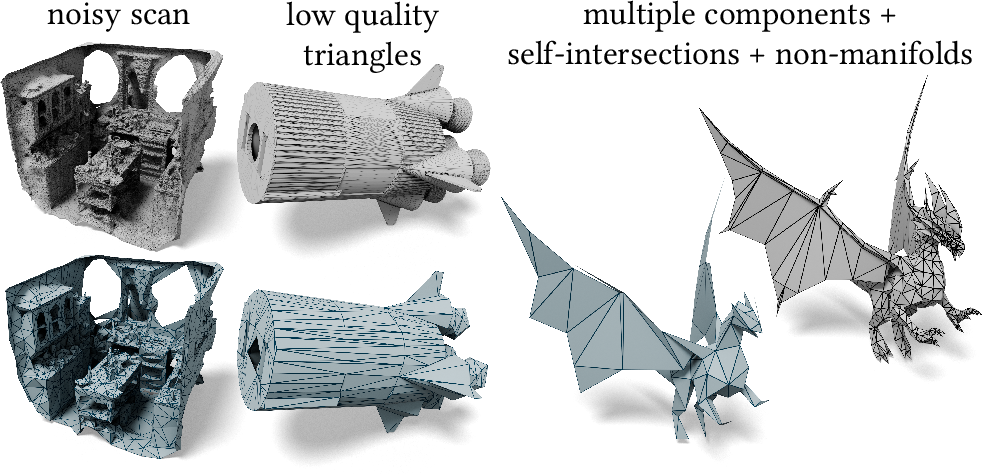}
    \end{center}
    \caption{\update{We stress test our method by simplifying a variety of challenging cases encountered in the wild, including noisy 3D scans (left), low-quality triangles from CAD models (middle), and meshes with defects (right). }}
    \label{fig:stress_tests}
\end{figure}
In this work, we propose a method to simplify \emph{any} textured triangle meshes that may contain multiple components, non-manifold geometry, and even the extreme case like triangle soups (\reffig{stress_tests}).
We elevate the manifold restriction of prior arts to work with a soup of triangles, while guaranteeing to converge to the classic quadric error simplification results on manifold inputs.
To handle textured meshes, we successively compute correspondences between the input mesh and the simplified mesh, unlike prior arts that aim at preserving UV coordinates. 
Our strategy reduces texture distortion and prevents texture bleeding occurs during the simplification process.
This combination enables us to handle arbitrary textured meshes one can encounter today, without degraded performance on clean inputs (see \reffig{manifold_vs_nonmanifold}).
We provide a background (\refsec{background}) and analyze the limitations of existing strategies (\refsec{pitfalls}). \update{This analysis highlights issues with current tools and motivates revisiting the simplification problem for mesh data in-the-wild.} 

\section{Related Work}\label{sec:mesh_simplification}
Our method belongs to the family of \emph{local} decimation schemes defined on triangle meshes, aiming to preserve their surface attributes. We thus focus our discussion on prior techniques closest to ours, and refer readers to \citet{luebke2003level} for a more comprehensive discussion.

\paragraph{Triangle Mesh Simplification}
Mesh simplification has been extensively studied in computer graphics to reduce the resolution while preserving the appearance of 3D objects. Early attempts tried to fit low resolution polygons \cite{DeHaemerZ91}, \emph{globally} remesh the input to a lower resolution \cite{HoppeDDMS92, Turk92}, or \emph{locally} remove each mesh element \cite{SchroederZL92}.
Since then, different formulations have been proposed, such as clustering of triangles \cite{CohenSteinerAD04, xu2024cwf} or vertices \cite{RossignacB93, LowT97, Lindstrom00}, greedily removing triangles \cite{Hamann94, GiengHJST98} or vertices \cite{Reinhard98}, simplifying a manifold envelope wrapping around the input \cite{MehraZLSGM09, Gao0P22, ChenPWVG23}, or optimization with a differentiable renderer \cite{HasselgrenMLAL21, deliot2024transforming, KnodtPWG24}.
Some of these techniques have also been revisited in the context of machine learning-driven simplification \cite{PotamiasPZ22, ChenKAJ23}.
Among them, perhaps the most popular strategy is the \emph{greedy edge collapse} \cite{Hoppe96} with the \emph{quadric error metric} \cite{GarlandH97}.
The popularity of edge collapses is partly due to the efficiency as each edge collapse exhibits constant time complexity.

\begin{figure}
    \begin{center}
    \includegraphics[width=1\linewidth]{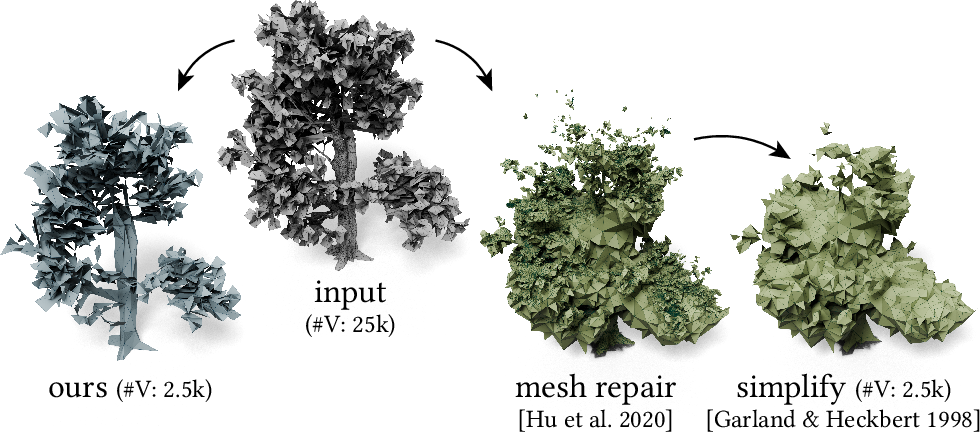}
    \end{center}
    \vspace{-7pt}
    \caption{\update{Repairing a triangle mesh with \cite{HuSWZP20} (third) and then simplifying the repaired model (fourth) depends on having a high quality repaired model, which is not always the case. Our method simplifies the input mesh directly and leads to a better result (first). }}
    \label{fig:tetwild_sim}
\end{figure}
\begin{figure}
    \begin{center}
    \includegraphics[width=1\linewidth]{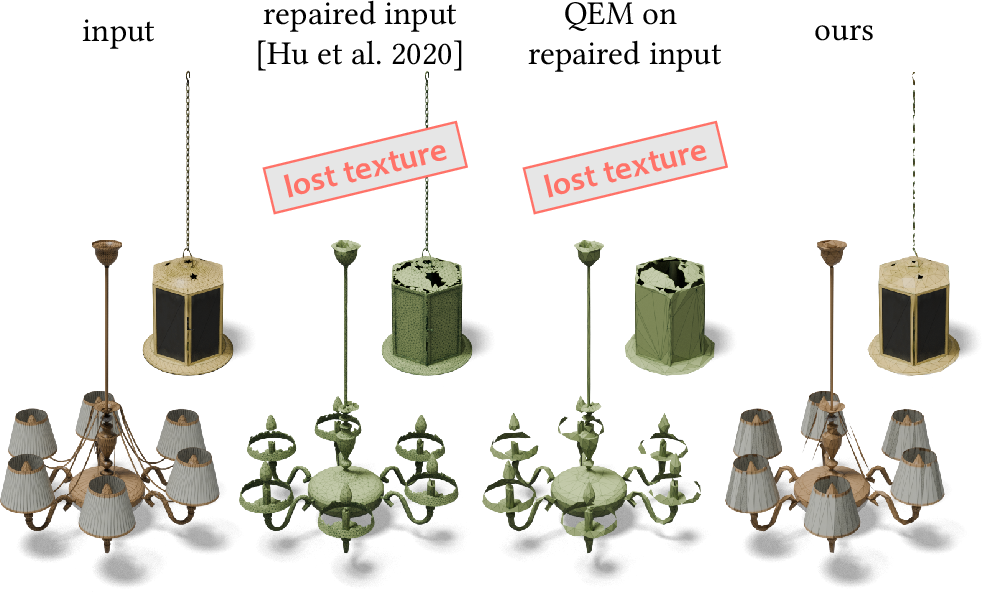}
    \end{center}
    \vspace{-7pt}
    \caption{Running mesh repairing \cite{HuSWZP20} (second column) and then simplifying the mesh with \cite{GarlandH97} (third column) leads to losses on surface attributes, such as textures, and may suffer from suboptimal results due to inperfect repairing. In contrast, our method directly simplifies the input, leading to better results (fourth column). }
    \label{fig:tetwild_textured_sim}
\end{figure}
\begin{figure}
    \begin{center}
    \includegraphics[width=1\linewidth]{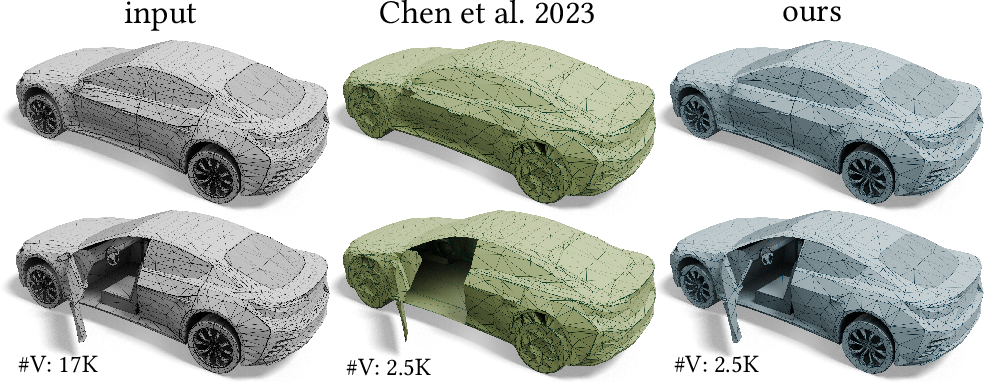}
    \end{center}
    \vspace{-7pt}
    \caption{\update{Comparing to methods that construct a manifold mesh wrapper and simplify it \cite{ChenPWVG23} (middle), our method directly simplifies the input and preserves interior structures (right).}}
    \label{fig:low_poly_car}
\end{figure}

\paragraph{Quadric Error Metrics}
The success of quadric error \cite{GarlandH97} has motivated a variety of \emph{quadrics} for measuring different geometric quantities. 
\citet{lindstrom1998fast} propose \emph{volume} and \emph{area} quadrics to measure the squared volume and area changes, respectively. 
\citet{QEMWithTexture} introduce the \emph{boundary quadric} to maintain the shape boundary. 
\citet{TrettnerK20} consider that a mesh element is a sample from a probability distribution and derive the \emph{probabilistic quadric} to further improve its numerical behavior.
Other extensions generalize the quadric error to incorporate surface attributes (e.g., textures) by defining quadric error in a higher dimension \cite{QEMWithTexture, GarlandZ05} or by linear approximation of the attribute \cite{NewQuadric}.
The optimal choice of metric varies depending on the input geometry and applications. The suite of quadric error metrics provides us with several options to mix-and-match them for different tasks, including the ones beyond mesh simplification such as quadric-inspired surface reconstruction \cite{KobbeltBSS01, JuLSW02, SchaeferW05} and filtering \cite{VieiraNMC10, LegrandTB19}.

\paragraph{Topology Varying Mesh Simplification}
Many of the methods mentioned above are designed for simplifying a single \emph{manifold} triangle mesh. Another class of mesh simplification generalizes the local decimation method to triangle meshes in the wild, which may contain multiple components or non-manifolds. 
\citet{GarlandH97} reformulate the edge collapse operation as an operation to collapse any \emph{vertex pair}, meaning one can merge two vertices even if they are not connected by an actual edge. These vertex pairs are computed based on vertex-to-vertex distance (see \reffig{v2v_distance}) and optimized with the quadric error metric.
\update{\citet{Schroeder97} identifies} different cases for manifold/non-manifold vertices and removes each vertex based on a distance-to-plane metric.
\citet{PopovicH97} strive for maximizing the compression rate and suggest not optimizing the vertex location.
In order to handle any triangle meshes, \citet{PopovicH97} reformulate it as a problem of simplifying \emph{simplicial complex}.
Our method is another instance of topology varying simplification for simplicial complex with high-quality geometric approximation and is capable of transferring textures.

Another possible direction is to \emph{repair} the mesh, i.e. convert it into a single manifold, and then simplify it. 
However, this depends on a repairing method to produce high-quality outputs, which is not always the case (see \reffig{tetwild_sim}) and can lose important surface attributes (see \reffig{tetwild_textured_sim}).
Alternatively, one can robustly create a manifold ``cage'' to wrap the triangle mesh and then simplify the manifold wrapper \cite{NooruddinT03, ChenPWVG23, PortaneriRHCA22}. However, these methods, e.g. \citet{ChenPWVG23}, often suffer from deleting interior components which may be important for interacting with the 3D object (see \reffig{low_poly_car}).
We thus follow the spirit of early attempts to simplify triangle meshes with vertex pair collapses.


\paragraph{Beyond Appearance Preservation}
Mesh simplification has also been studied in the context of preserving properties related to computation and simulation, such as preserving acoustic transfer \cite{LiFZ15}, spectral properties \cite{LiuJO19, ChenLJL20, KerosS23, LescoatLTJBO20}, intrinsic geometric quantities \cite{LiuGCSJC23, Shoemaker23}, and the dynamic behavior of physics-based simulations \cite{KharevychMOD09, Matusik15, ChenLMK17, ChenBWD018}.
When trained jointly with a machine learning model, one could even tailor-made a simplification for the underlying task, such as mesh classification \cite{HanockaHFGFC19, LudwigTC23}.
These methods serve as fundamental building blocks to obtain a coarse approximation for a simulation. 
But if one aims to build a multilevel solver that requires to transfer signals across resolutions \cite{GuskovSS99}, an important ingredient is to compute correspondences across the decimation hierarchy.
Such a motivation has stimulated several works on computing correspondences across hierarchy by computing an \emph{intrinsic} mapping for each local decimation \cite{LeeSSCD98, KhodakovskyLS03, LiuKCAJ20, LiuZBJ21, LiuGCSJC23} or by maintaining a bijective map with \emph{extrinsic} projection \cite{JiangSZP20, JiangZHSZP21}.
With such correspondences, one can then build hierarchical solvers for scalable computation \cite{LiuZBJ21, ZhangDFJJK22,WiersmaNEH23, Zhang23}.

\section{Pitfalls of Quadric Error Simplification}\label{sec:pitfalls}
Many existing simplification methods are designed for manifold meshes. Despite being a realistic assumption for meshes extracted from marching cubes, it becomes obsolete when considering artist-created assets in today's online repositories. The discrepancy often leads to geometric and texture issues.
\begin{figure}
    \begin{center}
    \includegraphics[width=1\linewidth]{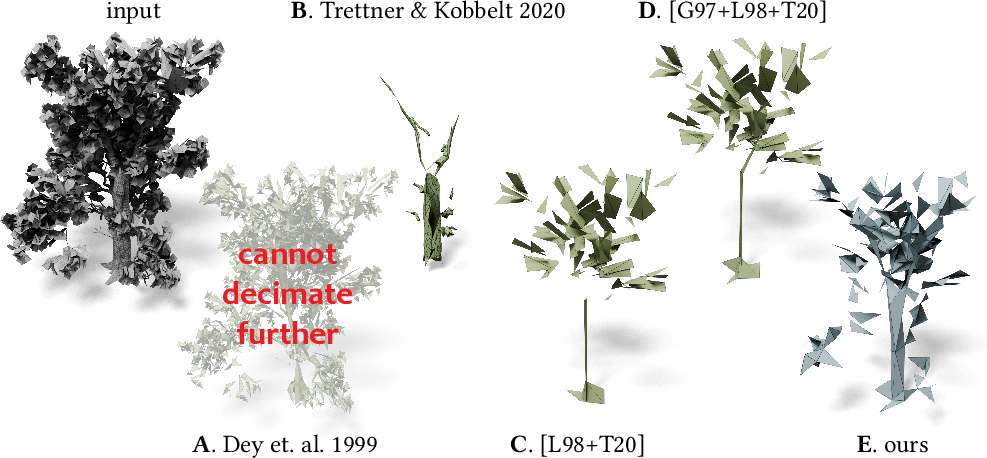}
    \end{center}
    \caption{Given a tree mesh with non-manifold elements and multi-connected components, we demonstrate that several off-the-shelf simplification methods and a combination of them suffer from different issues, such as failing at achieving target face counts or deleting planar components. In contrast, our method leads to perceptually better results when decimating the mesh down to 1\% of the original resolution.}
    \label{fig:pitfalls}
\end{figure}
\begin{figure}
    \begin{center}
    \includegraphics[width=1\linewidth]{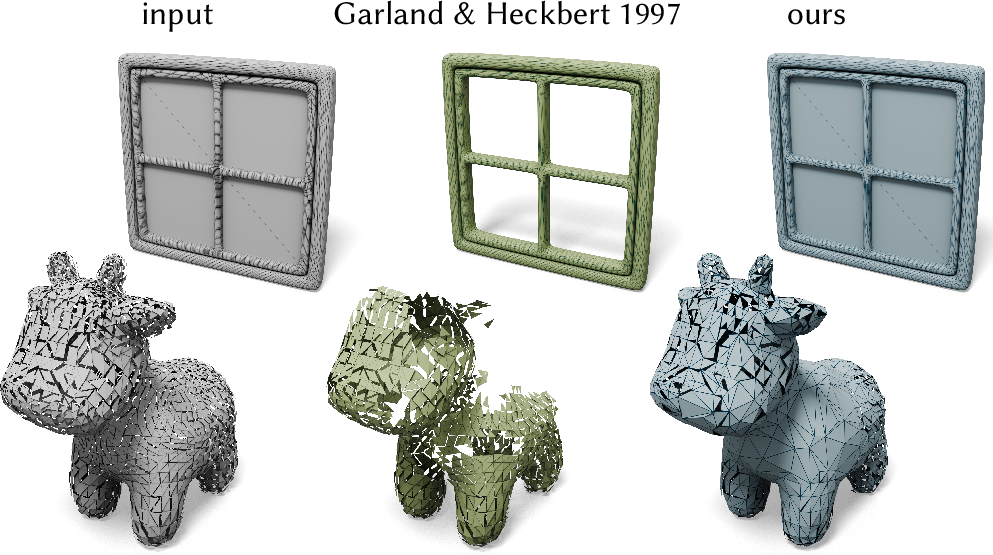}
    \end{center}
    \caption{\update{Simplifying the input mesh with the edge quadric error may lead to deleting geometrically significant parts because they lead to nearly zero quadric error (green). Our method overcomes such issues by penalizing collapses that will introduce significant area change (blue). }}
    \label{fig:qem_pitfalls}
\end{figure}

\subsection{Geometric Issues}
%
Many simplification methods, such as \cite{dey1999topology}, assume the input is a manifold mesh and aim at maintaining the topology throughout the simplification. However, this implicitly sets a lower bound on the coarsest mesh resolution because representing a component requires at least a single triangle (see \reffig{pitfalls} \textbf{A}).
Not to mention that many input meshes in practice already possess topological artifacts, which immediately break implementations that assume manifold inputs.
%

The edge quadric error metric \cite{GarlandH97, TrettnerK20} suffers from area losses. As described in \refsec{background}, quadric error measures the squared distance to triangle planes.  When a component is nearly \emph{developable}\update{, can be flattened to a plane without distortion}, quadric error will report zero error when collapsing edges in the developable region, resulting in the risk of removing large components and causing significant visual difference (see \reffig{qem_pitfalls}).
%

%
A potential remedy is to incorporate the \emph{area quadric} \cite{lindstrom1998fast}, in addition to the edge quadric error. 
Although it helps to preserve some large-area components, this combination (we use \LT to abbreviate the combination of area quadric \cite{lindstrom1998fast} and the probabilistic quadric \cite{TrettnerK20}) is still prone to deleting components due to no edges to collapse between disconnected parts.
%

%
\begin{wrapfigure}[6]{r}{0.83in}
    \vspace{-10pt}
	\includegraphics[width=\linewidth, trim={20mm 0mm 0mm 0mm}]{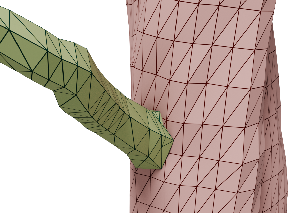}
	\label{fig:tree_parts}
\end{wrapfigure} 
In lieu of this, \citet{GarlandH97} suggest constructing virtual edges for vertex pairs such that their vertex-to-vertex distance is below a small threshold. 
However, in \reffig{pitfalls} \textbf{D}, we show that this strategy, a combination of \cite{TrettnerK20, lindstrom1998fast, GarlandH97} (called \GLT for short), still does not introduce significant improvements.
This is because computing distances between vertices is merely a sparse and inaccurate measure of distance between mesh components, leading to failures in constructing virtual edges between geometrically connected, but topologically disconnected components (see the inset).

\subsection{Texture Issues}\label{sec:texture_pitfalls}
\begin{figure}
    \begin{center}
    \includegraphics[width=1\linewidth]{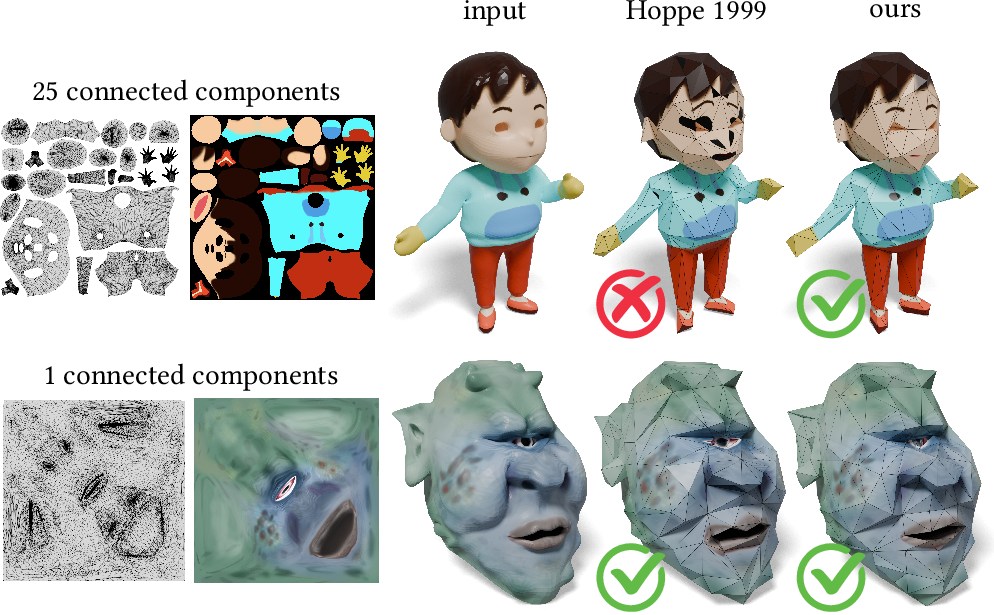}
    \end{center}
    \caption{Existing textured mesh simplification techniques work well when the input mesh has one or few connected components in the UV-space, a.k.a. UV islands (bottom middle). However, when deploying to meshes that have multiple UV islands, previous methods often lead to the ``texture bleeding'' artifact: seeing the background color of the texture image on the surface mesh (top middle). As most real-world meshes (e.g., \cite{MaggiordomoPCT20}) contain multiple UV islands, this motivates our method to handle both cases more robustly (right column). } 
    \label{fig:UV_islands}
\end{figure}
\begin{figure}
    \begin{center}
    \includegraphics[width=1\linewidth]{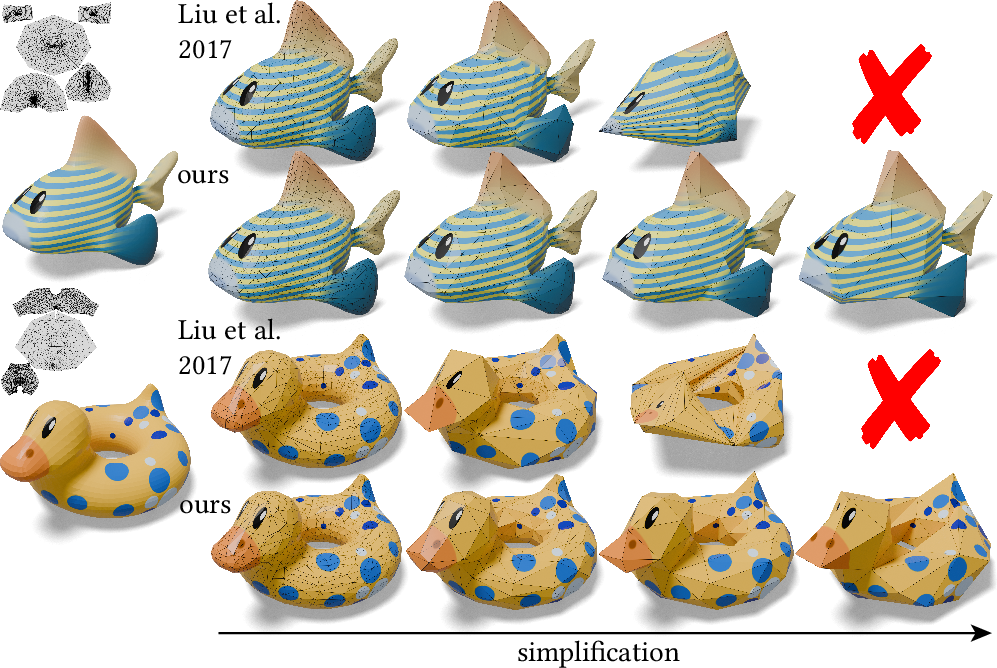}
    \end{center}
    \caption{An alternative method to avoid texture bleeding is to preserve the UV boundaries exactly, such as \cite{LiuFJG17}. But they often lead to larger geometric distortion and have a lower bound on the resolution of the simplified mesh (top rows). Our method, in contrast, can decimate the mesh aggressively while still being able to transfer texture colors.}
    \label{fig:seamless_comparison}
    \vspace{-5pt}
\end{figure} 
\begin{figure}
    \begin{center}
    \includegraphics[width=1\linewidth]{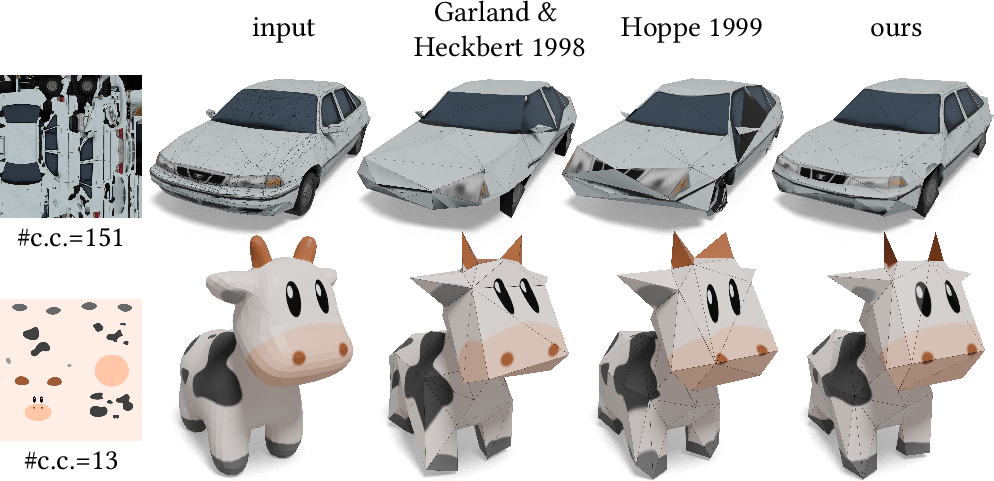}
    \end{center}
    \caption{Previous methods simplify a mesh while preserving its UV-coordinates for transferring texture to the simplified mesh. In addition to texture bleeding artifacts, these approaches can lead to distorted textures. Our method reduces such distortion by computing correspondences between the input and its simplified counterpart (right).}
    \label{fig:UV_preserving_decimation}
\end{figure}
\begin{figure}
    \begin{center}
    \includegraphics[width=1\linewidth]{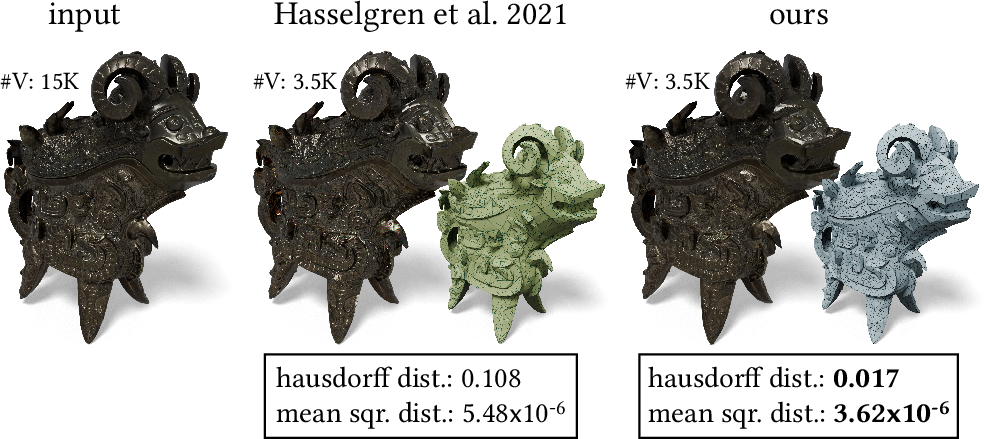}
    \end{center}
    \vspace{-3pt}
    \caption{Compared to an expensive mesh optimization method with differentiable rendering that usually takes a few hours \cite{HasselgrenMLAL21}, our method is orders of magnitude faster and achieves comparable quality.}
    \label{fig:appearance_driven}
\end{figure}

\emph{UV coordinates} are a popular way to store the mapping to a texture space that stores surface attributes. A traditional approach to preserve mesh attributes during simplification is to minimize geometric distortion of their UVs \cite{QEMWithTexture, NewQuadric}.
However, this strategy of \emph{approximately} preserving UV coordinates does not perform well when the input mesh has many connected UV components, a.k.a. UV islands (see \reffig{texture_quantitative}). 
This is because if the boundary of each UV island is not exactly preserved, and when the renderer interpolates texture colors near UV boundaries, distortion in the boundary curves could lead to interpolating the background color of the UV image onto the surface mesh. This behavior of rendering the background color on the mesh surface is also known as the ``texture bleeding'' artifact (see \reffig{UV_islands}).
In practice, a majority of the meshes have more than one UV islands, leading to significant texture bleeding (see \reffig{UV_islands}) or texture distortions (see \reffig{UV_preserving_decimation}) when applying this strategy to meshes in the wild. 
An alternative is to maintain the boundary shape of each UV island \emph{exactly} \cite{LiuFJG17}. However, this can lead to significant geometric distortion and possibly result in the early termination of the simplification process (see \reffig{seamless_comparison}).
On the other hand, although methods like differentiable rendering \cite{HasselgrenMLAL21} can potentially repair texture bleeding, such a texture optimization often leads to minutes to hours of additional computation time (see \reffig{appearance_driven}), making them unsuitable for real-time applications.
\section{Method}\label{sec:method}

In \refsec{pitfalls}, we highlight the pitfalls of the off-the-shelf methods (e.g., \cite{TrettnerK20, LiuFJG17}) failing at achieving satisfying results on wild meshes.
We further show that direct combinations of existing techniques, including \LT and \GLT suffer from various issues.
In this section, we present our method to address these issues in order to achieve high quality simplification on textured meshes in the wild.

\subsection{Vertex Pair Collapses}\label{sec:topological_operations}
\begin{wrapfigure}[6]{r}{1.1in}
    \vspace{-12pt}
	\includegraphics[width=\linewidth, trim={20mm 0mm 0mm 0mm}]{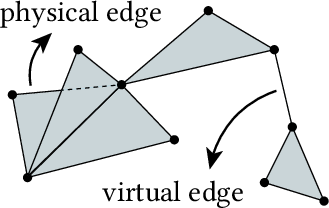}
	\label{fig:virtual_edges}
\end{wrapfigure}
Given a triangle mesh with vertices $V$ and faces $F$, we first construct a set of \emph{physical} and \emph{virtual} edges $E$ to connect vertex pairs.
We use \emph{physical edges} to denote the edges that are connected to at least one of the faces in $F$, and \emph{virtual edges} to denote the edges without neighboring faces (see inset). The combination $\M = (V,E,F)$, a.k.a. \emph{simplicial 2-complex}, is the input to our system.
\update{We decimate} the model by iteratively collapsing edges (1-simplices) in $E$ prioritized by our error metric in \refsec{error_metric}.

\subsubsection{Virtual Edges}\label{sec:virtual_edges}
\begin{wrapfigure}[7]{r}{1.1in}
    \vspace{-14pt}
	\includegraphics[width=\linewidth, trim={20mm 0mm 0mm 0mm}]{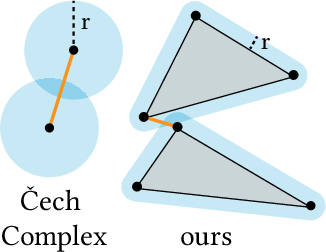}
	\label{fig:cech_complex}
\end{wrapfigure}
\begin{figure}
    \begin{center}
    \includegraphics[width=1\linewidth]{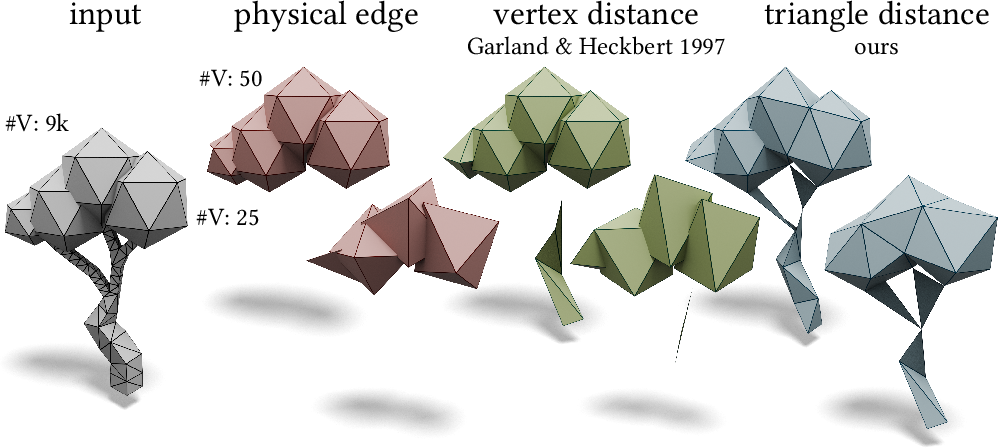}
    \end{center}
    \caption{\update{We simplify an input mesh (gray) with our method that computes virtual edges based on the triangle distances (blue). Our method can merge initially disconnected parts and lead to a more compact simplified mesh compared to the method by \citet{GarlandH97} which constructs virtual edges with vertex distances (green) and the method that is solely based on physical edges (red). }}
    \label{fig:v2v_distance_comparison}
\end{figure}
\begin{figure}
    \begin{center}
    \includegraphics[width=1\linewidth]{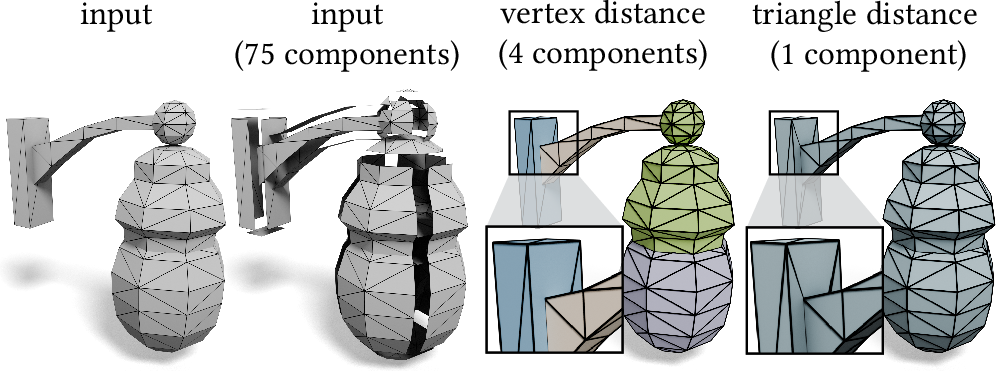}
    \end{center}
    \caption{\update{Given a lamp mesh representing a single connected rigid object (first), the raw data, though, may contain several disconnected pieces from the modeling process (second). Using vertex distance (third) to infer the connectivity still leads to 4 disconnected components indicated by different colors \cite{GarlandH97}. In contrast, using our triangle distance (fourth) leads to a single component. We visualize both physical and virtual edges with thick black lines in the third and fourth images.} }
    \label{fig:v2v_distance}
\end{figure}

Defining virtual edges is critical for decimating a wild mesh with multiple disconnected mesh components (see \reffig{v2v_distance_comparison}).
A common heuristic is to connect vertices with small vertex-to-vertex distance. However, it often leads to suboptimal results because topologically disconnected components that are geometrically close may still have large vertex-to-vertex distance due to discretization, such as \reffig{v2v_distance}.

We thus take inspiration from the construction of the \emph{\v{C}ech complex} to form virtual edges.
In a nutshell, \v{C}ech complex forms the connectivity between points if their $r$-radius balls have non-empty intersection.
In our case of triangle meshes, the natural generalization of the $r$-radius ball around a point becomes the $r$-offset surface around a triangle (see inset). We thus check connectivity between them by checking whether two offset surfaces have intersection. 
Intuitively, the construction suggests a straightforward implementation by computing triangle-to-triangle distances.
If the distance between two triangles is smaller than $2r$ \emph{and} the two triangles are not from the same connected component, we form a virtual edge between the two triangles by connecting their closest vertex pairs (see the inset).
Because our method additionally checks connected components, our result is \emph{not} a \v{C}ech complex, but it shares a similar spirit in the construction.

Once we have established an edge set $E$ with virtual edges and physical edges derived from $F$, we use a simplicial complex data structure to represent our input $(V,E,F)$ (detailed in \refsec{implementation}), then iteratively collapse edges (1-simplices) to decimate the mesh.
In \reffig{v2v_distance_comparison}, we demonstrate that our triangle distance measure is more effective than the naive vertex distance for simplifying meshes with multiple components.

\subsection{Error Metric}\label{sec:error_metric}
\begin{figure}
    \begin{center}
    \includegraphics[width=1\linewidth]{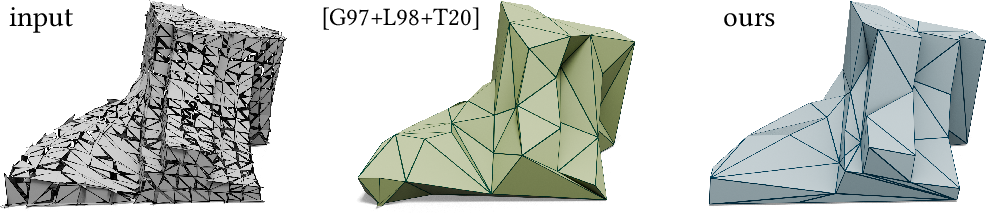}
    \end{center}
    \vspace{-5pt}
    \caption{Our novel quadric accumulation, the \emph{memory} edge quadric with the \emph{memoryless} area quadric, leads to a significantly better result in preserving sharp features compared to the straightforward combination of quadric \cite{GarlandH97} and area quadric \cite{lindstrom1998fast}. To illustrate the difference, we compare our method against such a combination, augmented with a tiny bit of probabilistic quadric \cite{TrettnerK20}) for numerical robustness and extended it with the data structure of \cite{PopovicH97} to handle non-manifolds. Given a soup of triangles (left), our method (right) leads to a better simplification result than the baseline (middle).  } 
    \label{fig:oversmooth}
    \vspace{-5pt}
\end{figure}

A key element in edge collapse algorithms is to define an error metric to prioritize the sequence of edge collapses. 
As pointed out in \refsec{pitfalls}, using the popular \emph{edge quadric error} \cite{GarlandH97} or its probabilistic version \cite{TrettnerK20} suffers from deleting large planar components. Naively adding the \emph{boundary quadric} \cite{QEMWithTexture} or the \emph{area quadric} \cite{lindstrom1998fast} may lead to oversmoothing (see \reffig{oversmooth}).
This motivates our modified metric to handle non-manifold meshes with multiple components. 

\begin{wrapfigure}[5]{r}{1.4in}
    \vspace{-14pt}
	\includegraphics[width=\linewidth, trim={20mm 0mm 0mm 0mm}]{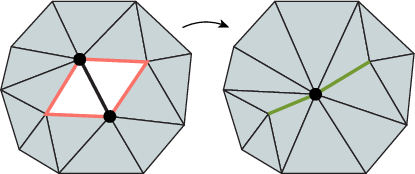}
	\label{fig:boundary_to_interior}
\end{wrapfigure}
Our key observation comes from studying the oversmoothing behavior of the area quadric.
When contracting a virtual edge, it can glue boundary edges (red edges in the inset) into interior edges (green edges in the inset). In such cases, the usual implementation of summing up quadrics (suggested in \cite{GarlandH97}) leads to area quadrics being applied to interior edges later in the decimation.
This behavior results in incorrectly high costs for removing interior edges. 
Collapsing an interior edge often leads to some area gain on one side of the edge and loss on the other. Although they should cancel each other and result in a small area change, area quadrics do not differentiate between the two (see \refsec{area_quadric}) and would simply square the area changes on both sides and sum up the results. 
This results in a high cost when decimating long interior edges and leads to overly uniform decimation.

To overcome the issue of oversmoothing, we propose a small, yet effective change in the quadric error metric accumulation. 
Specifically, we accumulate the edge quadric with the usual summation as proposed in \cite{GarlandH97}, a.k.a. the \emph{memory} implementation, but we do not accumulate the area quadric term, a.k.a. the \emph{memoryless} implementation. 
This targeted adjustment significantly improves sharp feature preservation when decimating a soup of triangles (see \reffig{oversmooth}). 
Crucially, our method ``converges'' to the standard quadric error metric \cite{GarlandH97} for closed manifold meshes. This ensures that our method maintains the excellent performance of \cite{GarlandH97} on manifold inputs, while improving the results for meshes in the wild.

\begin{figure}
    \begin{center}
    \includegraphics[width=1\linewidth]{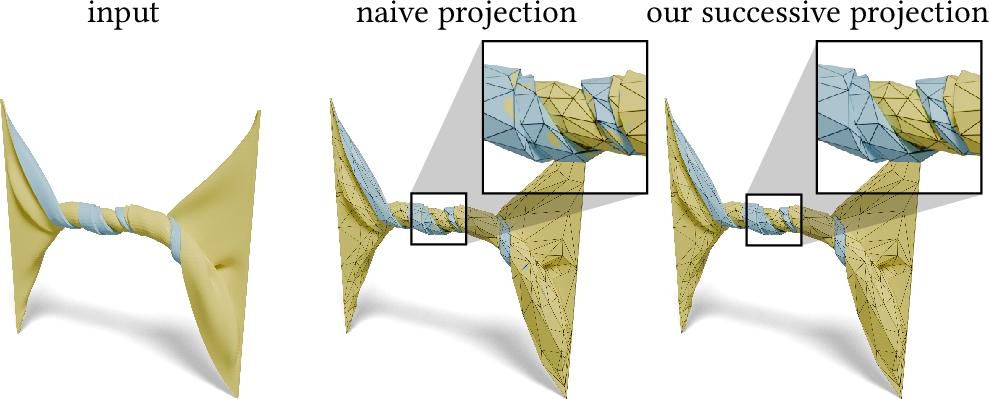}
    \end{center}
    \vspace{-7pt}
    \caption{On a thin shell cloth mesh with different colors (blue, yellow) on different sides, naive closest point projection between the simplified and the input meshes may lead to projecting points onto the wrong side (middle). In contrast, our successive closest point projection encourages the projection to lie on the corresponding side, leading to less projection error.}
    \label{fig:naive_projection} 
    \vspace{-5pt}
\end{figure}
\begin{figure}
    \begin{center}
    \includegraphics[width=1\linewidth]{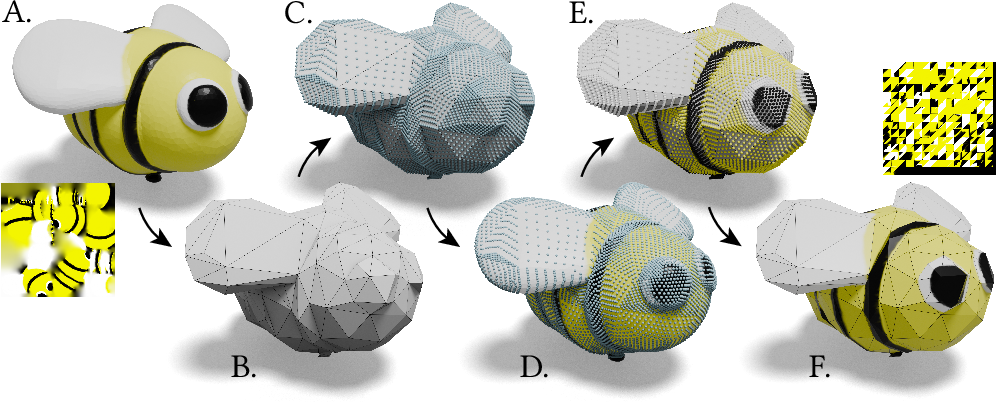}
    \end{center}
    \vspace{-7pt}
    \caption{Given an input mesh (A), we simplify it (B) while computing a successive map from the simplified mesh to the input. With this map, we can sample points on the simplified mesh (C), obtain correspondences to the texel locations, and transfer those points to the input with the correspondence map (D). Then we can simply look up the color information on the input (E) and generate a new texture image after simplification (F). }
    \label{fig:mesh_colors_process}
\end{figure}

\subsection{Texture Transfer With Successive Mapping}\label{sec:transfer_textures}
Previous textured mesh simplification methods preserve textures by preserving UV coordinates of the input mesh. This class of approach however suffers from issues detailed in \refsec{texture_pitfalls}.
We take inspiration from the idea of \emph{imposters} \cite{christiansen2005use} which bake a new texture to a simplified geometry via mapping. 
Our approach simplifies a mesh while keeping track of correspondences between the input mesh and its simplified counterpart with a \emph{successive mapping}, and then bake a new UV map after simplification. 
Our method avoids \emph{texture bleeding} (see \reffig{UV_islands}) and supports more aggressive simplifications (see \reffig{seamless_comparison}).

\subsubsection{Successive Projection}
\begin{wrapfigure}[5]{r}{1.3in}
    \vspace{-17pt}
	\includegraphics[width=\linewidth, trim={25mm 0mm 0mm 0mm}]{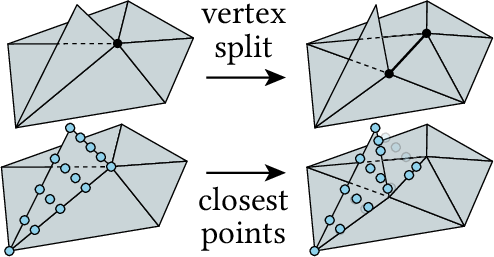}
	\label{fig:vertex_split}
\end{wrapfigure}
Given an input mesh $\M^0 = (V,E,F)$ and its coarsened counterpart $\M^c = (V_c, E_c, F_c)$, our goal is to compute a function $T: \M^c \rightarrow \M^0$ that maps a point $p \in \M^c$ on the simplified mesh to its corresponding point $T(p) \in \M^0$ on the input. Then, for any point $p$ on $\M^c$, we can retrieve attribute values from $T(p)$ and bake new texture maps for $M^c$.

Inspired by \cite{LiuKCAJ20}, we compute the mapping $T$ successively. Starting with an identity map from $\M^0$ to itself, we successively update the map during each edge collapse.
However, previous methods \cite{LiuKCAJ20} require the input to be a manifold mesh and without topological changes during simplification. This violates the scenario we considered, preventing us from using them.

We thus compute the mapping $T$ by successive closest point projection within the edge one-ring neighborhoods. 
Specifically, we store the decimation history $\{ \M^0, \M^1, \cdots, \M^c \}$ where two consecutive levels only differ by an edge collapse/vertex split. 
If a given point $p \in \M^c$ lies within the vertex one-ring of $\M^c$, we perform a vertex split to obtain $\M^{c-1}$, and project to the edge one-ring of $\M^{c-1}$ (see inset). We continue the local projection within the vertex one-ring until we reach $\M^0$. 
This localized projection encourages (but does not guarantee) projecting each point to the part of the surface it comes from, making it more robust to thin shell structures (see \reffig{naive_projection}).
%
%
One important implementation detail is that we reuse point locations on $\M^c$ for closest point queries, but using successive maps $\M^c \rightarrow \M^{c-1} \rightarrow \cdots \rightarrow \M^0$ to identify relevant edge one-rings is crucial to our results.
This ensures that the sampled surface attributes are geometrically closer to the simplified mesh $\M^c$ and avoids accumulating distortions from compositing multiple successive projections.

\begin{figure} 
    \begin{center}
    \includegraphics[width=1\linewidth]{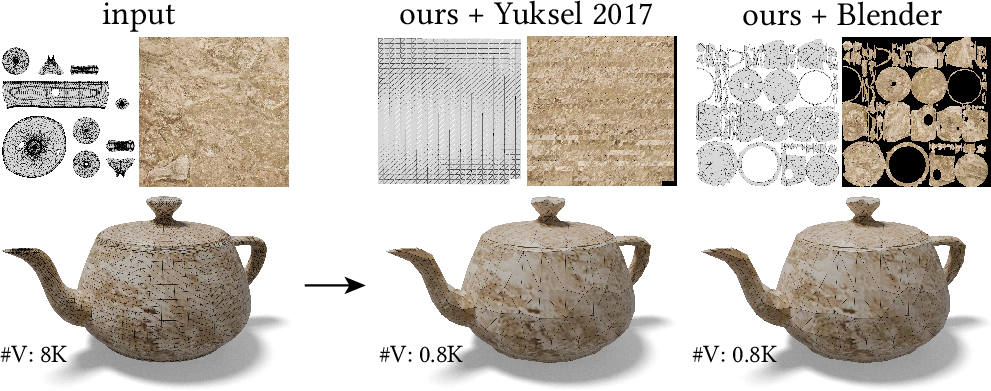}
    \end{center}
    \vspace{-7pt}
    \caption{\update{Our method is applicable to any injective texture mapping method, such as mesh color textures \cite{Yuksel17} and \textsc{Blender}'s smart UV. }}
    \label{fig:not_just_meshcolors}
\end{figure}

With the computed mapping $T$, we can bake any attributes from the input $\M^0$ to the simplified mesh $\M^c$ as textures via the process described in \reffig{mesh_colors_process}. Our mapping approach is compatible with any injective texture mapping technique (see \reffig{not_just_meshcolors}). In our paper, we use the \emph{mesh color texture} \cite{Yuksel17} due to its simplicity. 

\section{Results}\label{sec:results}
\begin{wrapfigure}[8]{r}{1.4in}
    \vspace{-10pt}
	\includegraphics[width=\linewidth, trim={25mm 0mm 0mm 0mm}]{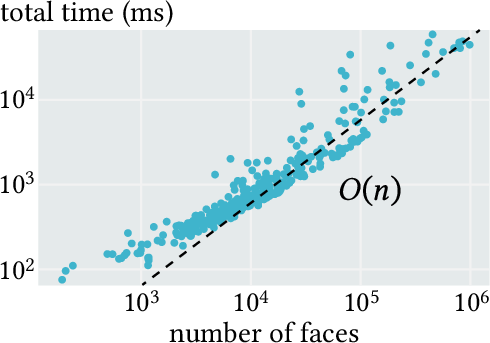}
	\label{fig:performance}
\end{wrapfigure}
Our method embrace the flexibility of existing QEM variants to control the decimation process, such as weighted simplification driven by visibility \cite{Hoppe97} (see \reffig{visibility_sim}).
We further enjoy the efficiency of being an edge collapse method that we can decimate hundreds of thousands of edges in a few seconds.
In the inset, we report the total runtime of decimating the meshes from the Polyhaven dataset down to 100 faces with around 56,000 texture sample points, evaluated on a MacBook with the M1 processor. 

\updatetwo{Our method shares the same asymptotic complexity, $O(n \log n)$, as previous edge collapse methods implemented with priority queues. However, our method has an extra precomputation of triangle-to-triangle distances for virtual edges, unlike previous methods which simply decimates existing physical edges. 
Compared to previous UV-preserving mesh simplification (e.g., \cite{QEMWithTexture}), our method is cheaper during simplification because we avoid the computation of high-dimensional quadrics for the UV coordinates. But we instead require an extra post-process to transfer textures. Specifically, our method requires to map each texel, represented as a barycentric point on the simplified mesh, to the input mesh by going through the decimation history. Luckily, this process is efficient because the barycentric coordinate for each texture sample will only be changed by a few edge collapses where the sample lies within the edge one-ring. This property allows us to skip computations for a majority of edge collapses when mapping texel points. From our implementation, computing correspondences for each point only involves less than 5 microseconds. 
When querying a large number of texels, this process can be trivially parallelized because the computation for each texel is independent.}

\bgroup
\def\arraystretch{1.5}
\renewcommand\tabularxcolumn[1]{m{#1}}
\newcolumntype{C}{>{\centering\let\newline\\\arraybackslash\hspace{0pt}}X}
\newcolumntype{L}{>{\raggedright\let\newline\\\arraybackslash\hspace{0pt}}X}
\begin{table}
    \setlength{\tabcolsep}{5.425pt}
    \centering
    \caption{
        We decimate the meshes of the Thingi10K dataset \cite{Thingi10K} down to 0.1\%/1\%/10\% of their original resolution and report the geometric errors (Hausdorff distance, mean squared Chamfer distance) averaged across the dataset. The meshes are normalized to have unit diagonal bounding boxes to avoid bias towards large meshes.
    }
    \begin{tabularx}{\linewidth}{>{\hsize=0.5\hsize}C | >{\hsize=0.26\hsize}C | >{\hsize=0.24\hsize}C}
    \toprule
        \emph{Methods} & \textit{Hausdorff}\text{\scriptsize $\times 10^{-1}$} & \textit{Chamfer}\text{\scriptsize $\times 10^{-1}$} \\
    \midrule
        \cite{QEMWithTexture} & 1.05/0.57/0.13 & 2.36/1.37/0.46   \\
        \cite{TrettnerK20} & 1.05/0.58/0.14 & 2.36/1.36/0.39  \\
    \rowcolor{derekTableBlue}
        Ours & 0.84/0.44/0.10 & 2.06/1.11/0.29 \\
    \bottomrule
    \end{tabularx}
    \label{tab:quantitative}
\end{table}
\bgroup
\bgroup
\def\arraystretch{1.5}
\renewcommand\tabularxcolumn[1]{m{#1}}
\newcolumntype{C}{>{\centering\let\newline\\\arraybackslash\hspace{0pt}}X}
\newcolumntype{L}{>{\raggedright\let\newline\\\arraybackslash\hspace{0pt}}X}
\begin{table}
    \setlength{\tabcolsep}{5.425pt}
    \centering
    \caption{
        We compute the symmetric Chamfer distance on the textures \cite{YuanKHMH18} for our method and the method by \citet{QEMWithTexture}. We decimate the mesh down to 1\% of the original resolution, and then report the average error from the \emph{Real-World Textured Things} \cite{MaggiordomoPCT20} and the Polyhaven datasets, respectively separated by ``/'', showing that our method quantitatively achieves lower texture errors.
    }
    \begin{tabularx}{\linewidth}{C|C}
    \toprule
        \emph{Methods} & \textit{Textured Chamfer}\text{\scriptsize $\times 10^{-1}$} \\
    \midrule
        \cite{QEMWithTexture} & 0.18 / 0.11  \\
    \rowcolor{derekTableBlue}
        Ours & 0.10 / 0.09 \\
    \bottomrule
    \end{tabularx}
    \label{tab:quantitative_texture}
\end{table}
\bgroup
\begin{figure}
    \begin{center}
    \includegraphics[width=1\linewidth]{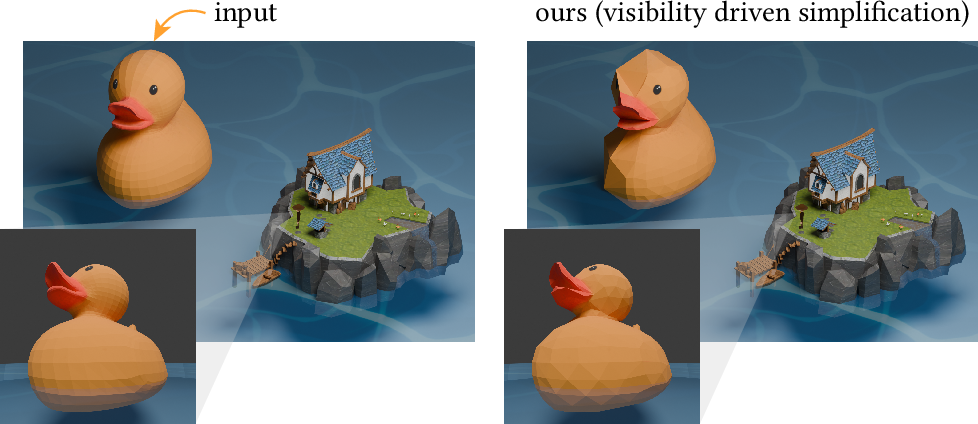}
    \end{center}
    \vspace{-7pt}
    \caption{\update{Our method is built on top of edge collapse algorithms, and thus can be seamlessly integrated with existing QEM extensions, such as simplifying a mesh driven by visibility \cite{Hoppe97}. In this example, we simplify the rubber duck statue based on the visible region (bottom left) from the observer on the island.}}
    \label{fig:visibility_sim}
\end{figure}
\begin{figure}
    \begin{center}
    \includegraphics[width=1\linewidth]{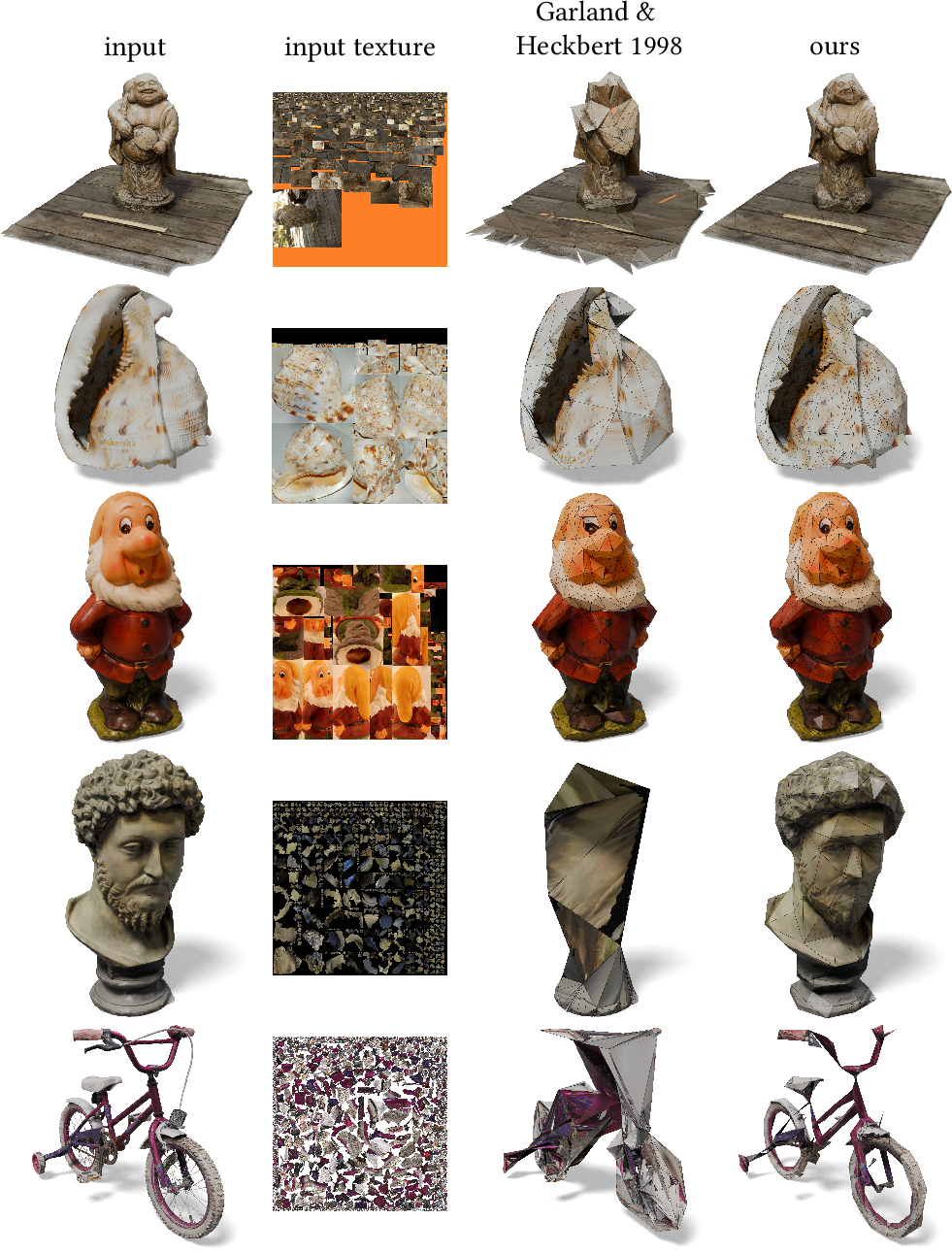}
    \end{center}
    \vspace{-7pt}
    \caption{We present a subset of textured mesh simplification results from \cite{QEMWithTexture} and ours on the \emph{Real-World Textured Things} dataset \cite{MaggiordomoPCT20}. A majority of them (the first column) have multiple texture islands (second column). Using the existing method (third column) suffers from texture bleeding and large distortion. In contrast, we obtain better results (fourth column) and avoid texture bleeding.}
    \label{fig:texture_quantitative}
\end{figure}

\paragraph{Qualitative Evaluations}
Our method offers several key advantages demonstrated throughout this paper. (1) Its virtual edge construction yields better results compared to previous method by \citet{GarlandH97} in \reffig{v2v_distance_comparison} and \reffig{pitfalls}. (2) Our enhanced error metric effectively prioritizes merging disconnected components while preventing the deletion of components with large surface areas (\reffig{manifold_vs_nonmanifold}, \reffignum{qem_pitfalls}). (3) The robust texture handling avoids texture bleeding, outperforming UV-preserving techniques (see \reffig{UV_preserving_decimation}), especially on non-manifold inputs with multiple components (\reffig{teaser}). Integrating these components, our system demonstrates quality improvements across numerous textured models in \reffig{polyhaven}, \reffignum{texture_quantitative}.
%


\paragraph{Quantitative Geometric Evaluations}
In \reftab{quantitative}, we quantitatively evaluate the geometric quality (excluding textures) of our simplification results against \cite{QEMWithTexture, TrettnerK20}, demonstrating that our method leads to smaller \emph{Hausdorff} and \emph{mean squared Chamfer} distances on the \emph{Thingi10K} dataset \cite{Thingi10K}. 
In \reffig{gallery} we show that our method can reliably decimate every mesh in the dataset down to 0.1\% of the original resolution for the entire dataset.

\paragraph{Quantitative Texture Evaluations}
We quantitatively evaluate textured mesh simplification on the \emph{Real-World Textured Things} \cite{MaggiordomoPCT20} and the Polyhaven datasets. Specifically, we decimate the meshes in the datasets that contain a single texture image (though potentially with multiple texture islands) down to 1\% of the input resolution. 
To avoid bias towards any specific render view, we measure the error using the average symmetric Chamfer distance on textures \cite{YuanKHMH18}, which measures the L2 norm of the color difference (color is normalized to [0,1]) between the closest spatial point pairs sampled on the surface (using ten thousands samples per mesh).
In \reftab{quantitative_texture}, our method consistently achieves lower errors than the method by \citet{QEMWithTexture}. We exclude the method by \cite{LiuFJG17} because it struggled to achieve target resolutions as shown in \reffig{polyhaven}.

\paragraph{User Studies}\label{sec:user_studies}
\begin{wrapfigure}[14]{r}{1.67in}
    \vspace{-15pt} 
	\includegraphics[width=\linewidth, trim={15mm 0mm 0mm 0mm}]{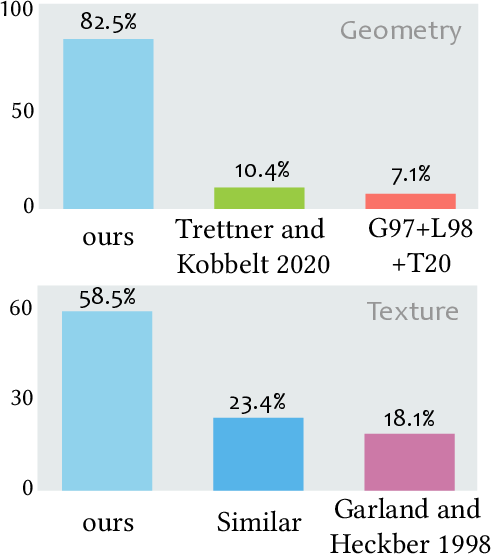}
	\label{fig:user_studies}
\end{wrapfigure}
In addition to the qualitative and quantitative evaluations, we performed user studies to get perceptual opinions on our mesh simplification results.
%
On the geometry side (see inset, top), we evaluated meshes presented in the \reffig{user_study_meshes}, showing that our outputs were preferred, in comparison to \cite{TrettnerK20} and the combination of baselines discussed in \refsec{pitfalls}. 
On the texture side, we conducted user study on randomly sampled meshes from the PolyHaven dataset, which contains a mixture of non-manifold textured meshes and manifold meshes extracted from isosurfacing. In the inset (bottom), we show that more than 80\% of the participants agree that our method is either comparable or outperforming \cite{QEMWithTexture}. Note that in this experiment, we further included an option to select \emph{Similar} quality in the study to verify the claim that our performance is comparable to \cite{QEMWithTexture} for manifold meshes.
We provide more details about \update{the studies} in \refsec{user_study_details}.

%

 

\section{Discussion \& Conclusion}
We have presented a practical mesh simplification method that can achieve better results on human-created meshes that commonly found online. We have also described an effective attribute transfer technique to preserve texture quality without impacting the quality of the mesh simplification process.
Our method extends the reach of mesh simplification to gracefully handle triangle meshes with inconvenient properties, including non-manifold meshes and meshes with multiple connected components.
This enhances performance by reducing the complexity of 3D models without compromising visual quality, enabling fast rendering and interactive simulation for real-time applications like online gaming, especially on low-power devices such as mobile phones and VR headsets, where computational resources are limited.

However, our current attribute transfer mechanism does not guarantee sampling from the outermost parts of the surface, potentially resulting in textures incorporating data from interior layers, which may be undesirable in certain cases (see \reffig{failure_case}). 
\begin{figure}
    \begin{center}
    \includegraphics[width=1\linewidth]{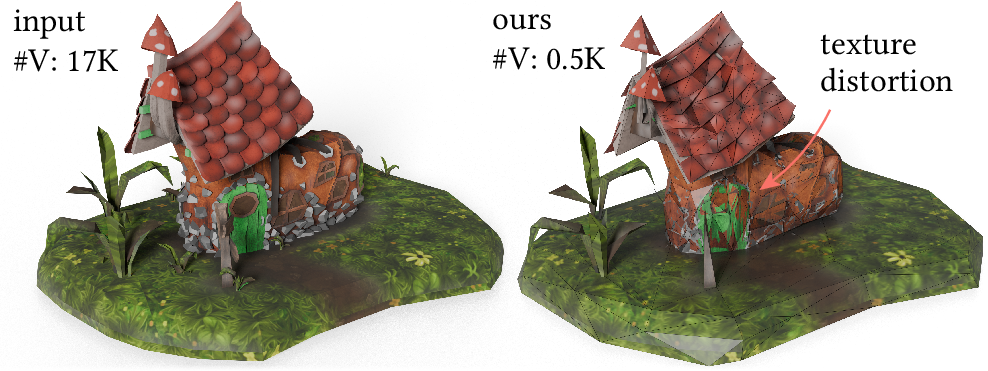}
    \end{center}
    \caption{\update{Our texture is based on a localized closest point projection. When the decimation introduces topological changes (e.g. deleting components), our projection may pick up colors from different components, leading to noticeable texture distortion.}}
    \label{fig:failure_case}
\end{figure}
Future work could address this problem by considering external visibility information while computing the mapping from the final mesh to the input surface.
%
\update{Employing adaptive texturing techniques based on the color content could mitigate potential blurry textures due to insufficient texel density in high-frequency areas. 
Exploring different stopping criteria, instead of simply setting target triangle counts, could further enhance the practical utility when deciding which geometric resolution to use for a given application.
Adding area quadrics reduces the risk of removing large area components, but future exploration to avoid deleting many small-area components (e.g., coniferous trees) could further enhance the visual perception on a wider variety of meshes.}
%
%
Our research contributes to the field of robust geometry processing and underscores the need to further enhance the robustness of downstream geometric algorithms, particularly because the output of our method, like the input, may contain defects such as non-manifold structures.
\begin{figure*}
    \begin{center}
    \includegraphics[width=1\linewidth]{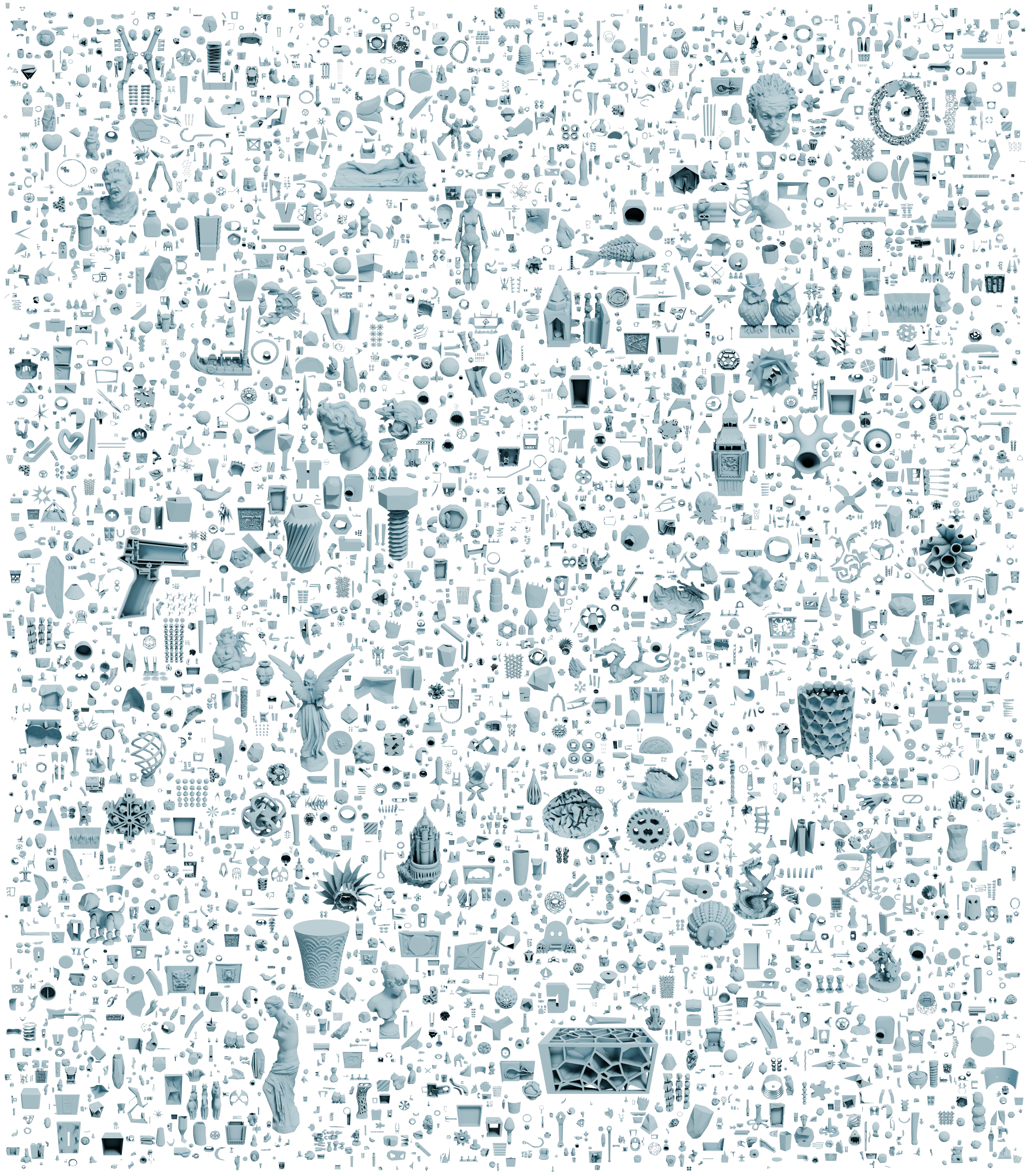}
    \end{center}
    \caption{\update{Our method can robustly decimation all 10 thousand meshes from \cite{Thingi10K} down to 1\% of their original resolution. Here we display a randomly selected subset of the simplified meshes.}}
    \label{fig:gallery}
\end{figure*} 
\begin{figure*}
    \begin{center}
    \includegraphics[width=1\linewidth]{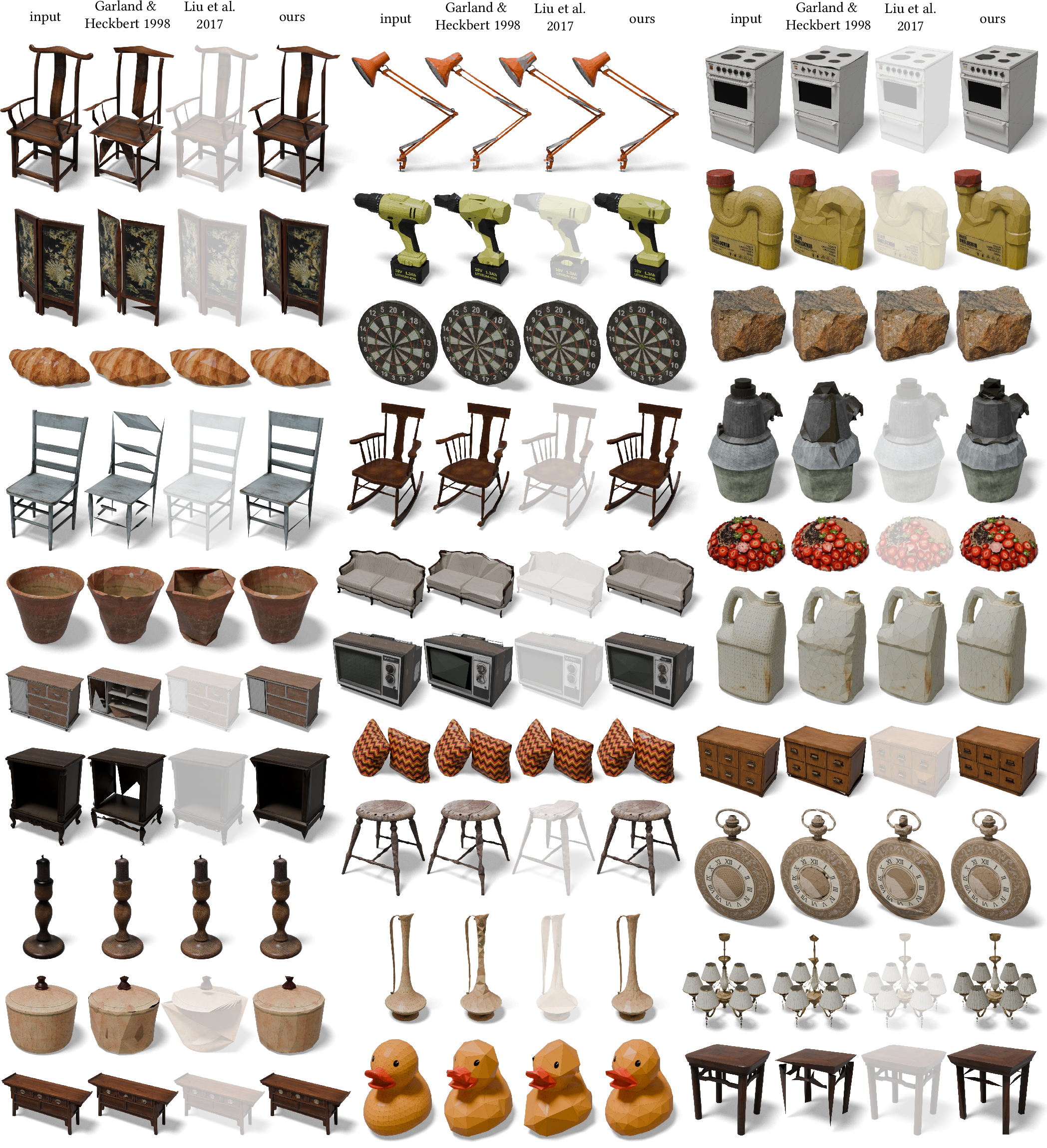}
    \end{center}
    \caption{\update{We show a subset of the textured mesh simplification results from the \emph{PolyHaven} dataset. We can observe several benefits of our method, such as preserving large planar regions, avoiding texture bleeding, and converging to comparable quality as previous methods when the input is a clean manifold mesh. We use faded colors to denote the failure in achieving target resolutions.}}
    \label{fig:polyhaven}
\end{figure*}
\bibliographystyle{ACM-Reference-Format}
\bibliography{sections/references}
\clearpage 
\appendix

\section{Background}\label{sec:background}
Our method simplifies a triangle mesh by iterative edge collapses. We prioritize the sequence of edge collapses by our modified \emph{quadric error metric}. Here, we briefly introduce the quadric error metric \cite{GarlandH97} and some variants that are relevant to our method.

\subsection{Quadric Error Metric}
%
Let $\vp \in \M$ be a point on a surface $\M$ embedded in $\R^3$ and $\n$ be the (unit) normal vector at $\vp$. The tangent plane at $\vp$ is given by all points $\vx \in \R^3$ that satisfy
\begin{align}\label{equ:plane_equation}
    \n^\top (\vx - \vp) = 0. 
\end{align}
The quadric error $E(\vx)$ measures the squared distance from any point $\vx$ to the tangent plane at $\vp$, which can be computed as 
\begin{align}\label{equ:quadric_error_metric}
    E(\vx) &= \big(\n^\top (\vx - \vp) \big)^2 \\
    & = (\vx - \vp)^\top \n \n^\top (\vx - \vp) \\
    & = \vx^\top \mA \vx + 2 \vb^\top \vx + c,
\end{align}
with
\begin{align}
    \mA = \n \n^\top, \quad \vb = -\mA \vp, \quad c = \vp^\top \mA \vp
\end{align}
A \emph{quadric} $Q$ refers to the triplet $(\mA, \vb, c)$, which are quantities derived by the plane equation $\P$ with a unit normal $\n$ and a point $\vp$ on the tangent plane:
\begin{align}\label{equ:quadric}
    Q \coloneq (\mA, \vb, c) = \P(\n, \vp),
\end{align}
This quadric $Q$ gives us the complete information to compute the quadric error $E(\vx)$. 
%

\paragraph{Triangle Quadrics}
The quadric error can be generalized to triangulated 2-manifolds by defining the quadric error $E_{ijk}$ on the plane of each triangle $ijk$ 
\begin{align}
    E_{ijk}(\vx)  = \vx^\top \mA_{ijk} \vx + 2 \vb_{ijk}^\top \vx + c_{ijk}.
\end{align}
This defines the \emph{triangle quadric} $Q_{ijk}$ as
\begin{align}\label{equ:triangle_quadrics}
    Q_{ijk} = (\mA_{ijk}, \vb_{ijk}, c_{ijk}) = \P(\vn_{ijk}, \vv_i)
\end{align}
derived from the face normal $\n_{ijk}$ and the location of a point on the plane, such as one of the triangle corner vertices $\vv_i$.

\paragraph{Vertex Quadrics}
For each vertex $i$, the \emph{vertex quadric error} $E_i$ is defined as the weighted summation of the triangle quadric errors $E_{ijk}$ from its one-ring triangles $ijk$
\begin{align}\label{equ:vertex_quadric_error}
    E_i(\vx) &= \sum_{ijk \in \mathcal{N}_i} a^i_{ijk} E_{ijk}(\vx) \\
    & = \sum_{ijk \in \mathcal{N}_i} a^i_{ijk} (\vx^\top \mA_{ijk} \vx + 2 \vb_{ijk}^\top \vx + c_{ijk}) \\
    & = \vx^\top \Big( \smashoperator{\sum_{ijk \in \mathcal{N}_i}} a^i_{ijk} \mA_{ijk} \Big) \vx + 2 \Big( \smashoperator{\sum_{ijk \in \mathcal{N}_i}} a^i_{ijk} \vb_{ijk} ^\top\Big) \vx +  \Big( \smashoperator{\sum_{ijk \in \mathcal{N}_i}} a^i_{ijk} c_{ijk} \Big) \nonumber
\end{align}
\begin{wrapfigure}[5]{r}{0.66in}
	\includegraphics[width=\linewidth, trim={20mm 0mm 0mm 0mm}]{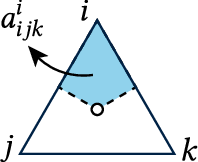}
	\label{fig:barycentric_vertex_area}
\end{wrapfigure}
We use $\mathcal{N}_i$ to denote the one-ring triangles of the vertex $i$ and $a^i_{ijk}$ is a portion of the vertex area at $i$ coming from face $ijk$.
A common choice of area is the \emph{barycentric area} (see inset) which simply sets $a^i_{ijk} = a_{ijk} / 3$ to be one-third of the face area $a_{ijk}$ \cite{MeyerDSB02}.
This derivation gives rise to the definition of \emph{vertex quadric} $Q_{i}$ as a weighted summation of its one-ring triangle quadrics $Q_{ijk}$
\begin{align}
    Q_i = \sum_{ijk \in \mathcal{N}_i}\!\! a^i_{ijk} Q_{ijk}
\end{align}
where the summation between quadrics is the component-wise summation for all elements in the triplets. 

\paragraph{Edge Quadrics}
The quadric error metric $E_{ij}$ for each edge $ij$ is defined as the summation of vertex quadric errors of its endpoints 
\begin{align}\label{equ:edge_quadric_error}
    E_{ij}(\vx) = E_i(\vx) + E_j(\vx)
\end{align}
As we can see in \refequ{vertex_quadric_error} that summing up quadric errors leads to a summation of quadrics, this leads to the definition of an \emph{edge quadric} $Q_{ij}$ as
\begin{align}\label{equ:edge_quadric}
    Q_{ij} = Q_i + Q_j.
\end{align}

\subsection{Quadric Error Edge Collapses}\label{sec:background_edge_collapses}
\citet{GarlandH97} suggest simplifying a triangle mesh by iteratively collapsing the edge with the smallest edge quadric error $E_{ij}(\vx^\star)$ which is the value of $E_{ij}$ evaluated at the optimal location $\vx^\star$ that minimizes $E_{ij}$.
Specifically, given an edge $ij$, edge collapse is performed by replacing this edge with a new vertex $i'$ that is located at $\vx^\star$.
%
%
Let $Q_{ij} = (\mA_{ij} , \vb_{ij}, c_{ij})$, the optimal location $\vx^\star$ can be computed solving a linear system obtained from setting $\nabla E_{ij} = 0$
\begin{align}\label{equ:optimal_location}
    \mA_{ij} \vx^\star = -\vb_{ij}
\end{align}
If the matrix $\mA_{ij}$ is invertible and well-conditioned, one can solve for $\vx^\star$ with standard linear solvers, such as Cholesky decomposition. If not, one can use the singular value decomposition to solve for $\vx^\star$ more robustly \cite{Lindstrom00} or use the probabilistic quadric as a regularization \cite{TrettnerK20}.

In subsequent iterations, instead of recomputing the edge quadric with \refequ{edge_quadric}, \citet{GarlandH97} recommend using the $Q_{ij}$ as the vertex quadric for the newly inserted vertex. Suppose we collapse an edge $ij$ to a vertex $i'$, the quadric of the new vertex $Q_{i'} = Q_{ij}$ is simply the edge quadric before the collapse, and $Q_{i'}$ will then be used to update the quadrics for its one-ring edges.
This definition allows the quadric error to ``memorize'' all the plane information on the \emph{input} mesh, instead of from the current mesh. In practice, compared to recomputing the edge quadric $Q_{ij}$ from scratch, such an accumulation often leads to a more efficient and desirable decimation. But some applications may favor not accumulating quadrics \cite{NewQuadric}.

We use the term ``memory'' to denote the implementation that accumulates the quadric error metric throughout the decimation and ``memoryless'' to denote the implementation that recomputes the quadric at every iteration.  

\subsection{Area Quadrics}\label{sec:area_quadric}
The quadric error metric has stimulated several variants to serve different purposes (see \refsec{mesh_simplification}). Among them, the \emph{area quadric} is related to our decimation metric.
The area quadric measures the \emph{squared area change} for each edge collapse. 
\citet{lindstrom1998fast} show that the squared area change for each edge $ij$ can be written as a quadratic expression, thus giving birth to the area quadric $\Qa_{ij}$ 
\begin{align}\label{equ:area_quadric}
    \Qa_{ij} = \sum_{ab \in \mathcal{N}_{ij} \cap \partial \M}\!\! \Big([\vs_{ab}]_\times^\top [\vs_{ab}]_\times,\ -[\vs_{ab}]_\times \vt_{ab},\ \vt_{ab}^\top \vt_{ab} \Big) / 2
\end{align}
with
\begin{align}
    \vs_{ab} = \vv_b - \vv_a \qquad \vt_{ab} = \vv_a \times \vv_b
\end{align}
where $ab$ are $ij$'s one-ring edges that are also on the mesh boundary $\partial \M$ and $[\cdot]_\times$ is the \emph{cross product matrix} defined as
\begin{align}
    [\vx]_\times \coloneq 
    \begin{bmatrix}
        0 & -x_2 & x_1 \\
        x_2 & 0 & -x_0 \\
        -x_1 & x_0 & 0
    \end{bmatrix}
\end{align}
\begin{wrapfigure}[5]{r}{1.4in}
    \vspace{-12pt}
	\includegraphics[width=\linewidth, trim={15mm 0mm 0mm 0mm}]{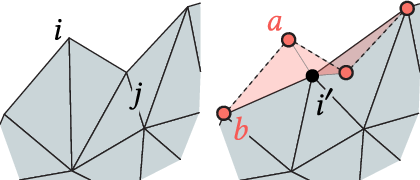}
	\label{fig:area_quadric}
\end{wrapfigure}
This area quadric $\Qa_{ij}$ is derived from summing up the squared area (computed with cross products) from each boundary edge $ab$ to the newly inserted vertex $i'$ (see inset). This implies that this area quadric is merely an approximation to the actual area change because this measure ignores the cancelation between the area gain and the loss. 
But this proxy maintains the favorable quadratic expression and can be seamlessly incorporated into the quadric-based mesh simplification.

\section{Theoretical Guarantees}
\paragraph{No Texture Bleeding} 
{\update{Our method guarantees no texture bleeding. This is because each texel (represented as a barycentric point) on the simplified mesh is guaranteed to be projected onto a face of the input mesh. This ensures that no texel will pick up the background color of the input texture, thus avoiding texture bleeding.

\paragraph{Manifold In, Manifold Out} 
Our simplification framework offers flexibility regarding topological changes. The vanilla configuration allows topological changes, such as merging components, to strive for high quality simplification (see \reffig{polyhaven}).
Alternatively, for applications requiring strict topology preservation on known manifold inputs, our method can be augmented with validity checks (e.g., those detailed in Appendix C of \cite{LiuKCAJ20}), to reject edges that violate them.
Including these checks ensures that our decimation maintains the manifold structure.

\paragraph{Convergence to QEM on Closed Manifolds}
Our error metric converges to the edge quadric error metric \cite{GarlandH97} when the input mesh is a closed manifold mesh. 
This property is guaranteed because we only augment the area quadric if the local edge one-ring contains boundary edges (as discussed in \refsec{error_metric}). Thus, for a closed manifold mesh without boundary edges, no area quadric will be included in the decimation, and our method converges to QEM up to some minor implementation differences, such as how to handle tied quadric error. 
This design decision ensures that our method improves LOD on meshes in the wild, without sacrificing the already amazing performance of QEM on manifolds.}}
%
%

\section{Implementation}\label{sec:implementation}
In addition to the proposed improvements over the edge topology, the error metric, and the texture-preserving simplification part, reproducing our method requires changes in the implementation (e.g., data structures) to properly handle non-manifold meshes. 
Here, we focus on discussing the differences in implementation compared to the standard approach, such as in \cite{GarlandH97}, and omit the aspects that remain the same (e.g. using a priority queue to prioritize edge costs, employing a linear solve to determine the optimal vertex positions, etc.).

\subsection{Simplicial Complex Data Structure}
\begin{wrapfigure}[7]{r}{1.4in}
    \vspace{-10pt}
	\includegraphics[width=\linewidth, trim={25mm 0mm 0mm 0mm}]{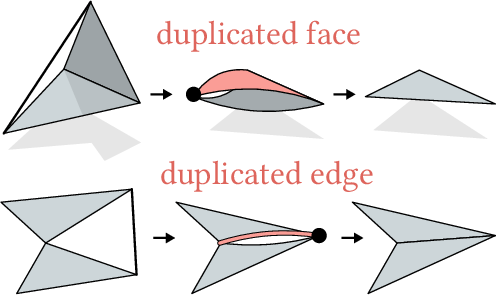}
	\label{fig:degenerated_simplicies}
\end{wrapfigure}
Many off-the-shelf mesh simplification methods are implemented using the \emph{half-edge data structure} \cite{MullerP78}. However, such a data structure is primarily designed to handle manifold meshes with implicit assumptions e.g. each edge is shared by one or two faces, making it inapplicable to defective meshes.

Several data structures for manipulating simplicial complexes have been proposed to handle general simplices (e.g., high dimensional simplices) and reduce the memory footprint \cite{FlorianiH05}.
However, in our case, these more generic data structures are unnecessary. We only perform edge collapses on simplicial 2-complexes. Thus, instead of using general-purpose data structures, we use a list of lists (similar to \cite{PopovicH97}) to represent the simplicial complex.
Specifically, we construct three lists to store the \emph{star} information for each (1) vertex to its one-ring edges, (2) vertex to its one-ring faces, and (3) edge to its one-ring faces. 
For each edge collapse, we use these lists to gather neighboring simplices and update their connectivity correspondingly.
If a collapse leads to duplicated or degenerated simplices (see inset), we simply remove them from the corresponding lists.

\subsection{Successive Texture Transfer}
{\update{To support the texture transfer described in \refsec{transfer_textures}, for each edge collapse, we additionally store the local edge one-ring mesh before the collapse and the local vertex one-ring mesh after the collapse for closest point successive projection. This is because successive projections only requires updating texture samples within the one-ring neighborhoods, we can avoid redundant computations by skipping samples outside the one-ring (by checking the corresponding face of each barycentric point) and only perform the closest point projection within the one-ring meshes. In practice, the number of triangles within the one-ring is small, we did not observe significant efficiency improvement if one build spatial hierarchies for the local one-ring meshes. }}

\section{User Study Details}\label{sec:user_study_details}
We performed two user studies to gather perceptual opinions on our mesh simplification results. Specifically, the participants were instructed to perform the user study within the context of a common online game scenario where the players are using low-end devices (e.g., mobile phones) to play online video games with other players. Due to the high performance requirements (e.g., frame rates, bandwidth), the 3D assets in the game have to satisfy a fixed budget (e.g., the total number of vertices in the scene). Under this situation, we asked participants to fill out online surveys about their preferences among a variety of simplified 3D assets obtained from our method and baselines.

One of the studies evaluated the quality of textured mesh simplification. For each response, we randomly selected a mesh from the PolyHaven dataset, simplified with our method and the textured QEM by \citet{QEMWithTexture}, and asked the participant to select their preferences among (1) our method, (2) \cite{QEMWithTexture}, and (3) comparable (see \reffig{polyhaven} for some examples in the dataset). 
We included \emph{comparable} as one of the options because our method ``converges'' to QEM when the input is a closed manifold mesh and this can avoid users randomly picking one out of comparable outputs. 
The results from the two methods are presented in an arbitrary order in the survey to avoid the order bias.
From the 591 responses we collected, 81.9 \% of them indicated that our method is either comparable or outperforming the baseline, suggesting that our method leads to better simplification results (see \refsec{results} for more details). 

The second study solely focused on the quality of geometry (excluding textures) after simplification. We presented results from the method by \cite{TrettnerK20} and our invented baseline \GLT mentioned in \refsec{pitfalls}. 
Participants were asked to select their favorite mesh simplification results out of the 10 user-created non-manifold meshes presented in the paper (see \reffig{user_study_meshes} for the collection) representing a wide range of meshes from organic shapes, 3D scenes, and man-made objects. 
From the total 690 responses (69$\times$10), 82.5\% of them favored our results, 10.4\% favored the results from \cite{TrettnerK20}, and 7.1\% favored [G97+L98+T20] (see \refsec{results}).
\begin{figure}
    \begin{center}
    \includegraphics[width=1\linewidth]{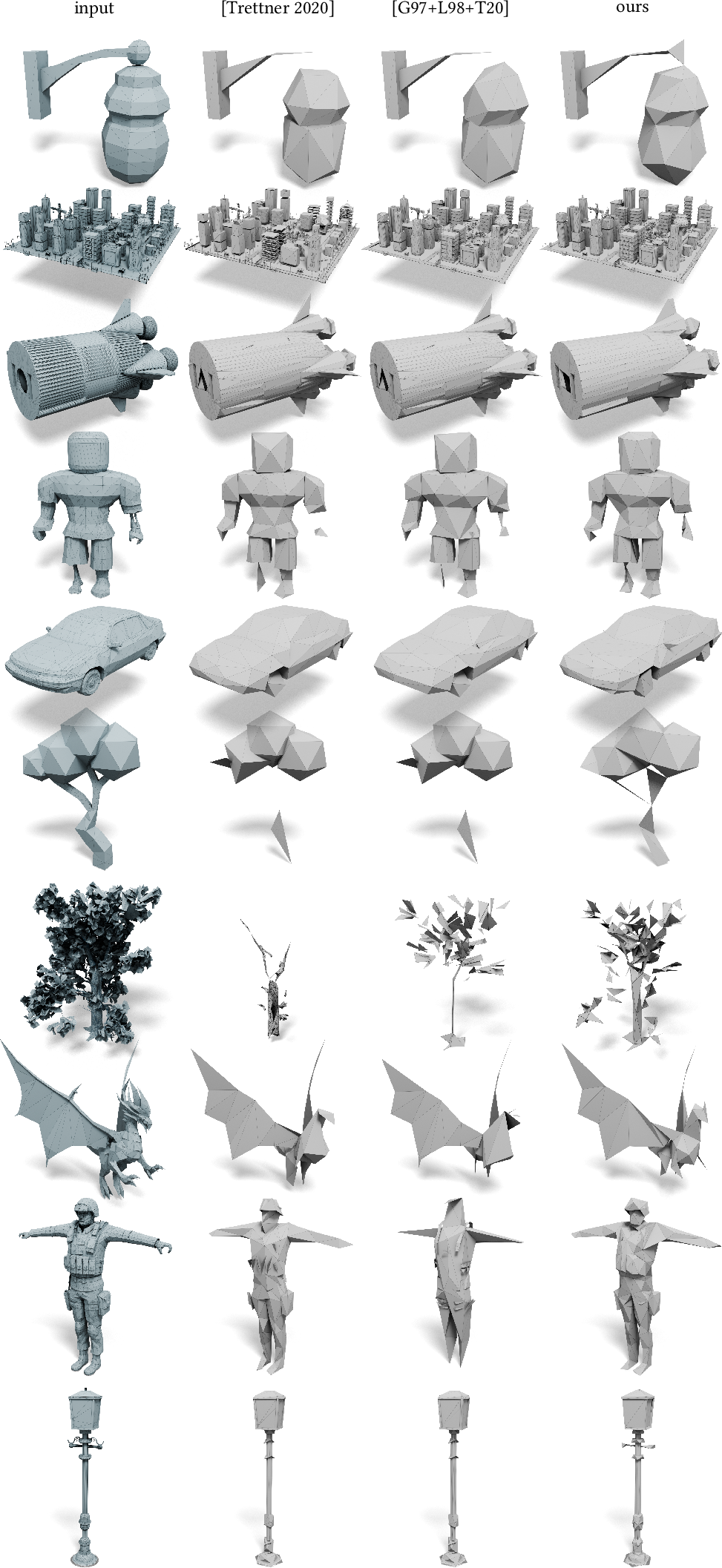}
    \end{center}
    \caption{We conducted a user study to evaluate the perceptual preference across different mesh simplification techniques, including (1) ours, (2) \cite{TrettnerK20}, and (3) \GLT a combination of \cite{TrettnerK20, GarlandH97, lindstrom1998fast}. 
    The figure contains all the meshes used in the user studies. Note that in the user study, the order of different methods is randomly permuted to avoid order bias, but in this figure we display them in a consistent order for clarify.}
    \label{fig:user_study_meshes}
\end{figure}

For these two studies, we invited 73 participants (51 male, 21 female, 1 non-binary individual, age ranging from 11-50, 75\% graduate students and 25\% working professionals including engineers and artists) to conduct the user study.

\end{document}